\documentclass[11pt,reqno]{amsart}
\usepackage{amsmath,amsthm,amssymb,mathrsfs,stmaryrd}
\usepackage[all]{xy}
\usepackage{url}
\usepackage{geometry}
\usepackage[dvips]{color}
\usepackage{amsxtra}
\usepackage{amscd}
\usepackage{eucal}
\usepackage{epsfig}
\usepackage{graphics}

\usepackage{bm}

\usepackage[utf8]{inputenc}
\usepackage[OT1]{fontenc}

\usepackage{relsize}
\usepackage[bbgreekl]{mathbbol}
\usepackage{amsfonts}
\usepackage{textgreek}

\DeclareSymbolFontAlphabet{\mathbb}{AMSb}
\DeclareSymbolFontAlphabet{\mathbbl}{bbold}

\renewcommand{\bold}{\mathbf}
\newcommand{\RR}{\mathbb{R}}

\newcommand{\R}{\mathbb{R}}

\newcommand{\Q}{\mathbb{Q}}
\newcommand{\Z}{\mathbb{Z}}

\newcommand{\g}{\mathfrak{g}}

\newcommand{\nc}{\newcommand}
\nc{\on}{\operatorname}
\nc{\la}{\lambda}
\nc{\wh}{\widehat}
\nc{\wt}{\widetilde}
\nc{\sw}{{\mathfrak s}{\mathfrak l}}
\nc{\ghat}{\wh{\g}}
\nc{\hhat}{\wh{\h}}
\nc{\mc}{\mathcal}
\nc{\bi}{\bibitem}
\nc{\pa}{\partial}
\nc{\ppart}{(\!(t)\!)}
\nc{\pparl}{(\!(\la)\!)}
\nc{\zpart}{(\!(z^{-1})\!)}
\nc{\n}{{\mathfrak n}}
\nc{\ol}{\overline}
\nc{\mb}{\mathbf}
\nc{\bb}{{\mathfrak b}}
\nc{\su}{\wh\sw_2}
\nc{\h}{{\mathfrak h}}
\nc{\can}{\on{can}}
\nc{\ntil}{\wt{\n}}
\nc{\pone}{{\mathbb P}^1}
\nc{\bs}{\backslash}
\nc{\al}{\alpha}
\nc{\gt}{{\mathfrak g}'}
\nc{\ds}{\displaystyle}
\nc{\ep}{\varepsilon}
\nc{\alp}{\alpha}
\nc{\nab}{\nabla}
\nc{\RN}{\RR^n}
\nc{\RM}{\RR^m}
\nc{\Sc}{\mathrm{Schr}\ddot{\mathrm{o}}\mathrm{dinger}}
\nc{\Scd}{Schr\ddot{o}dinger}
\nc{\hb}{\hbar}
\nc{\st}{\section}
\nc{\sst}{\subsection}
\nc{\ta}{\theta}
\nc{\mm}{\mathrm}
\nc{\fr}{\frac}
\nc{\nb}{\nabla}
\nc{\pt}{\partial}
\nc{\ra}{\rightarrow}
\nc{\La}{\Leftarrow}
\nc{\Ra}{\Rightarrow}
\nc{\LRa}{\Leftrightarrow}
\nc{\lra}{\leftrightarrow}
\nc{\tim}{\times}
\nc{\oli}{\overline}
\nc{\uli}{\underline}
\nc{\ch}{\check}
\nc{\lt}{\left}
\nc{\rt}{\right}
\nc{\lap}{\Delta}
\nc{\infi}{\infty}
\nc{\intf}{\int_{-\infty}^\infty}
\nc{\de}{\delta}
\nc{\Gm}{\Gamma}
\nc{\gm}{\gamma}
\nc{\sse}{\subset}
\nc{\ssen}{\subsetneq}
\nc{\nsse}{\not\subset}
\nc{\lds}{\ldots}
\nc{\cds}{\cdots}
\nc{\cdt}{\cdot}
\nc{\dds}{\ddots}
\nc{\vds}{\vdots}
\nc{\tet}{\text}
\nc{\mr}{\mapsto}
\nc{\ml}{\mapsfrom}
\nc{\di}{\mm{dist}}
\nc{\til}{\tilde}
\nc{\ba}{\bigcap}
\nc{\bu}{\bigcup}
\nc{\com}{\circ}
\nc{\ssim}{\approx}
\nc{\sq}{\sqrt}
\nc{\rou}{\rho}
\nc{\tf}{\textbf}
\nc{\se}{\simeq}
\nc{\iden}{\mathbb{1}}
\nc{\disu}{\amalg}
\nc{\inclu}{\hookrightarrow}
\nc{\xra}{\xrightarrow}
\nc{\FF}{\mathbb{F}}
\nc{\KK}{\mathbb{K}}
\nc{\NP}{\mathbb{N}_+}
\nc{\emp}{\varnothing}
\nc{\sB}{\mathscr{B}}
\nc{\sR}{\mathscr{R}}
\nc{\Lm}{\Lambda}
\nc{\sm}{\setminus}
\nc{\rra}{\rightrightarrows}
\nc{\op}{\oplus}
\nc{\bop}{\bigoplus}
\nc{\ssum}{\Sigma}

\nc{\diam}{\mm{diam}}
\nc{\diag}{\mm{diag}}
\nc{\msr}{\mathscr}
\nc{\lan}{\langle}
\nc{\ran}{\rangle}
\nc{\rua}{\rightharpoonup}
\nc{\rp}{\RR\mm{P}}
\nc{\PP}{\mathbb{P}}
\nc{\af}{\mathbb{A}}
\usepackage{tikz-cd}
\usetikzlibrary{cd}
\usetikzlibrary{decorations.pathmorphing}
\usepackage{tkz-euclide}
\usetikzlibrary{patterns}
\usetikzlibrary{shapes}
\usetikzlibrary{shapes.misc}
\usetikzlibrary{calc,math}
\usepackage{ifthen}
\usetikzlibrary{calc}
\tikzset{
  my label/.style={font=\scriptsize,inner sep=2pt},
  a/.style={my label,above,node contents={$a$}},
  b/.style={my label,right,node contents={$b$}},
  a-1/.style={my label,above,node contents={$a^{-1}$}},
  b-1/.style={my label,right,node contents={$b^{-1}$}},
}
\newcommand\caley[6]{
  \ifthenelse{0<#1}{
    \pgfmathtruncatemacro\newlev{#1-1}
    \pgfmathtruncatemacro\len{#2}
    \draw[draw=black,-latex] (0,0) -- (\len pt,0) node[pos=.6,#3] coordinate (O);
    \begin{scope}[shift={(O)}]
      \begin{scope}[rotate=90] \caley{\newlev}{\len/2}{#4}{#5}{#6}{#3} \end{scope}
      \begin{scope}[rotate=0]  \caley{\newlev}{\len/2}{#3}{#4}{#5}{#6} \end{scope}
      \begin{scope}[rotate=-90]\caley{\newlev}{\len/2}{#6}{#3}{#4}{#5} \end{scope}
    \end{scope}
  }{\fill[black] circle(1pt);}
}
\usetikzlibrary{lindenmayersystems,arrows.meta}
\newcount\quadrant
\pgfdeclarelindenmayersystem{cayley}{
  \rule{A -> B [ R [A] [+A] [-A] ]}
  \symbol{R}{ \pgflsystemstep=0.5\pgflsystemstep } 
  \symbol{-}{
    \pgfmathsetcount\quadrant{Mod(\quadrant+1,4)}
    \tikzset{rotate=90}
  }
  \symbol{+}{
    \pgfmathsetcount\quadrant{Mod(\quadrant-1,4)}
    \tikzset{rotate=-90}
  }
  \symbol{B}{
    \draw [dot-cayley] (0,0) -- (\pgflsystemstep,0) 
       node [font=\footnotesize, midway, 
         anchor={270-mod(\the\quadrant,2)*90}, inner sep=.5ex] 
           {\ifcase\quadrant$a$\or$b$\or$a^{-1}$\or$b^{-1}$\fi};
    \tikzset{xshift=\pgflsystemstep}
  }
}
\tikzset{
  dot/.tip={Circle[sep=-1.5pt,length=3pt]}, cayley/.tip={Stealth[]dot[]}
}
\nc{\hl}{\colorbox{yellow}}
\nc{\tcr}{\textcolor{red}}
\nc{\sA}{\mathscr{A}}
\nc{\sC}{\mathscr{C}}
\nc{\sD}{\mathscr{D}}
\nc{\sS}{\mathscr{S}}
\nc{\sT}{\mathscr{T}}
\nc{\id}{\mm{id}}
\nc{\Id}{\mm{Id}}
\nc{\opp}{\mm{o}}
\nc{\Ob}{\mm{Ob}}

\nc{\Ad}{\mm{Ad}}
\nc{\uAd}{\underline{\mathrm{Ad}}}
\nc{\wad}{/\uAd}
\nc{\ad}{\mm{ad}}
\nc{\Aut}{\mm{Aut}}
\nc{\Inv}{\mm{Inv}}
\nc{\Gal}{\mm{Gal}}
\nc{\sig}{\sigma}
\nc{\sgn}{\mm{sgn}}
\nc{\Sym}{\mm{Sym}}
\nc{\sym}{\mm{sym}}
\nc{\vphi}{\varphi}
\nc{\vi}{\varphi}
\nc{\vp}{\varpi}
\nc{\vP}{\varPi}
\nc{\bF}{\mathbf{F}}
\nc{\fT}{\mathfrak{T}}
\nc{\fN}{\mathfrak{N}}

\nc{\kap}{\kappa}
\nc{\fX}{\mathfrak{X}}
\nc{\fY}{\mathfrak{Y}}

\nc{\fa}{\mathfrak{a}}
\nc{\fb}{\mathfrak{b}}
\nc{\fB}{\mathfrak{B}}
\nc{\bss}{\mathcal{B}}
\nc{\opn}{\mm{Op}}
\nc{\ops}{\mm{op}}
\nc{\ff}{\mathcal{F}}
\nc{\res}{\tet{res}}
\nc{\OO}{\mathcal{O}}
\nc{\fm}{\mathfrak{m}}
\nc{\fM}{\mathfrak{M}}
\nc{\fn}{\mathfrak{n}}
\nc{\fp}{\mathfrak{p}}
\nc{\fq}{\mathfrak{q}}
\nc{\Mor}{\tet{Mor}}
\nc{\aff}{\mathbb{A}}
\nc{\grass}{\tet{Grass}}
\nc{\sur}{\twoheadrightarrow}

\usepackage{stmaryrd}
\nc{\llb}{\llbracket}
\nc{\rrb}{\rrbracket}
\nc{\cok}{\mm{Coker}}
\nc{\fU}{\mathfrak{U}}
\nc{\wtil}{\widetilde}
\nc{\Ann}{\mathrm{Ann}}
\nc{\fc}{\mathfrak{c}}
\nc{\Ver}{\mm{Ver}}

\nc{\GL}{\mm{GL}}
\nc{\gl}{\mathfrak{gl}}
\nc{\SO}{\mm{SO}}
\nc{\so}{\mathfrak{so}}
\nc{\SL}{\mm{SL}}
\nc{\fsl}{\mathfrak{sl}}
\nc{\mO}{\mm{O}}
\nc{\Bil}{\mm{Bil}}
\nc{\Alt}{\mm{Alt}}

\nc{\gr}{\mm{gr}}
\nc{\End}{\mm{End}}
\nc{\Idem}{\mm{Idem}}
\nc{\lto}{\leadsto}
\nc{\tor}{\mm{Tor}}
\nc{\ext}{\mm{Ext}}
\nc{\lext}{\uli{\mm{Ext}}}
\nc{\loc}{\mm{loc}}
\nc{\Loc}{\mm{Loc}}
\nc{\pr}{\mm{pr}}
\nc{\cO}{\mathcal{O}}
\nc{\cI}{\mathcal{I}}
\nc{\cA}{\mathcal{A}}
\nc{\cD}{\mathcal{D}}
\nc{\cU}{\mathcal{U}}
\nc{\disc}{\mm{disc}}
\nc{\cls}{\mm{cls}}
\nc{\gen}{\mm{gen}}
\nc{\rep}{\mm{rep}}
\nc{\ade}{\mathbb{A}}
\nc{\pslr}{\mm{PSL}_2(\R)}
\nc{\pslz}{\mm{PSL}_2(\Z)}
\nc{\slr}{\mm{SL}_2(\R)}
\nc{\sor}{\mm{SO}_2(\R)}
\nc{\stab}{\mm{Stab}}
\nc{\acts}{\curvearrowright}
\nc{\tang}{\mm{T}^1\mathbb{H}}
\nc{\HH}{\mathbb{H}}
\nc{\x}{\times}
\nc{\xx}{^\times}
\nc{\xs}{^*}
\nc{\xp}{^\perp}
\nc{\Frac}{\mm{Frac}}
\newcommand{\pp}{%
  \mathrel{\ooalign{$\lneq$\cr\raise.22ex\hbox{$\lhd$}\cr}}}
\nc{\ie}{i.e.~}
\nc{\resp}{resp.~}
\nc{\eg}{e.g.~}
\nc{\cf}{cf.~}
\nc{\tiff}{if and only if}
\nc{\enu}{enumerate}
\nc{\hangin}{\hangindent\leftmargini\textup{(i)}~}
\nc{\hangma}{\hspace*{\leftmargini}}
\nc{\indnull}{\mathbb{o}}
\nc{\inv}{^{-1}}
\nc{\du}{^\vee}
\nc{\sups}{\supset}
\nc{\Hom}{\mm{Hom}}
\nc{\iHom}{\mathcal{H}\!\mathit{om}}
\nc{\mat}{\mm{Mat}}
\nc{\Set}{\mm{Set}}
\nc{\Rng}{\mm{Rng}}
\nc{\Grp}{\mm{Grp}}
\nc{\abGrp}{\mm{AbGrp}}
\nc{\LGrp}{\mm{LieGrp}}
\nc{\CLGrp}{\mm{ConnLieGrp}}
\nc{\CSCLGrp}{\mm{CSCLieGrp}}
\nc{\LAlg}{\mm{LieAlg}}
\nc{\TGrp}{\mm{TopGrp}}
\nc{\Diff}{\mm{Diff}}
\nc{\Deck}{\mm{Deck}}
\nc{\Fld}{\mm{Fld}}
\nc{\Top}{\mm{Top}}
\nc{\Ab}{\mm{Ab}}
\nc{\Mod}{\tet{-}\mm{Mod}}
\nc{\Alg}{\tet{-}\mm{Alg}}
\nc{\kFld}{\tet{-}\mm{Fld}}
\nc{\GG}{\mathbb{G}}
\nc{\Sch}{\mm{Sch}}
\nc{\sch}{\mm{sch}}
\nc{\lrs}{\mm{lrs}}
\nc{\red}{\mm{red}}
\nc{\Fun}{\mm{Fun}}
\nc{\sing}{\mm{sing}}
\nc{\dr}{\mm{dR}}
\nc{\G}{\mathbb{G}}
\nc{\bt}{\bullet}
\nc{\bqp}{\oli{\Q}_p}
\nc{\qp}{\Q_p}
\nc{\zp}{\Z_p}

\newcommand{\thrm}[1]{Theorem~\ref{#1}}

\nc{\dasha}{\dasharrow}
\nc{\Etale}{\'Etale}
\nc{\etale}{\'etale}

\nc{\nil}{\mathfrak{N}}
\nc{\jac}{\mathfrak{J}}
\nc{\rad}{\mm{rad}}
\nc{\lrad}{\mm{l\tet{-}rad}}
\nc{\rrad}{\mm{r\tet{-}rad}}
\nc{\inj}{\rightarrowtail}

\nc{\psupp}{\tet{-supp}}

\nc{\ot}{\otimes}
\nc{\bx}{\wh{\ot}}
\nc{\ha}{\hat{H}}
\nc{\cE}{\mathcal{E}}
\nc{\cF}{\mathcal{F}}
\nc{\cG}{\mathcal{G}}
\nc{\cH}{\mathcal{H}}
\nc{\cL}{\mathcal{L}}
\nc{\cB}{\mathcal{B}}
\nc{\dlim}{\varinjlim}
\nc{\ilim}{\varprojlim}
\nc{\sU}{\mathscr{U}}
\nc{\sV}{\mathscr{V}}
\nc{\sW}{\mathscr{W}}
\nc{\chH}{\check{H}}
\nc{\sI}{\mathcal{I}}
\nc{\psh}{\mm{PSh}}
\nc{\sh}{\mm{Sh}}
\nc{\sep}{\mm{sep}}
\nc{\alg}{\mm{alg}}
\nc{\ab}{\mm{ab}}
\nc{\setmid}{\,|\,}
\nc{\nsetmid}{\!\nmid\!}
\nc{\smid}{\,|\,}
\nc{\bmid}{\,\big|\,}
\nc{\Bmid}{\,\Big|\,}
\nc{\CP}{\mathbb{C}\mm{P}}
\nc{\RP}{\mathbb{R}\mm{P}}
\nc{\HP}{\mathbb{H}\mm{P}}
\nc{\FP}{\mathbb{F}\mm{P}}
\nc{\ltri}{\triangleleft}
\nc{\tri}{\triangle}
\nc{\act}{\curvearrowright}
\nc{\Isom}{\mm{Isom}}
\nc{\Om}{\Omega}
\nc{\om}{\omega}
\nc{\ev}{\mm{ev}}
\nc{\std}{\mm{std}}
\nc{\triv}{\mm{triv}}
\nc{\sob}{\mm{sob}}

\nc{\lip}[1]{\left\| #1 \right\|_\mm{Lip}}
\nc{\cl}{\mm{cl}}
\nc{\Cl}{\mm{Cl}}
\usepackage{upgreek}
\nc{\pres}{\prescript}
\nc{\ttau}{\pres{t}{}\tau}
\nc{\ttl}[1]{\pres{t}{}\tau^{\le#1}}
\nc{\ttg}[1]{\pres{t}{}\tau^{\ge#1}}
\nc{\tl}[1]{\tau^{\le#1}}
\nc{\tg}[1]{\tau^{\ge#1}}
\nc{\trl}{\sT^{\le0}}
\nc{\trll}[1]{\sT^{\le#1}}
\nc{\trg}{\sT^{\ge0}}
\nc{\trgg}[1]{\sT^{\ge#1}}
\nc{\tH}{\pres{t}{}H}
\nc{\tF}{\pres{t}{}F}
\nc{\sr}[1]{#1^\sharp}
\nc{\Fr}{\mm{Fr}}

\makeatletter
\newcommand{\xinclusur}[2][]{%
  \lhook\joinrel
  \ext@arrow 0359\rightarrowfill@ {#1}{#2}%
  \mathrel{\mspace{-15mu}}\rightarrow
}
\makeatother

\makeatletter
\newcommand{\xlinclusur}[2][]{%
  \lhook\joinrel
  \ext@arrow 0359\rightarrowfill@ {#1}{#2}%
  \mathrel{\mspace{-23.25mu}}\leftarrow
}
\makeatother

\makeatletter
\newcommand*{\relrelbarsep}{.386ex}
\newcommand*{\relrelbar}{%
  \mathrel{%
    \mathpalette\@relrelbar\relrelbarsep
  }%
}
\newcommand*{\@relrelbar}[2]{%
  \raise#2\hbox to 0pt{$\m@th#1\relbar$\hss}%
  \lower#2\hbox{$\m@th#1\relbar$}%
}
\providecommand*{\rightrightarrowsfill@}{%
  \arrowfill@\relrelbar\relrelbar\rightrightarrows
}
\providecommand*{\leftleftarrowsfill@}{%
  \arrowfill@\leftleftarrows\relrelbar\relrelbar
}
\providecommand*{\xrightrightarrows}[2][]{%
  \ext@arrow 0359\rightrightarrowsfill@{#1}{#2}%
}
\providecommand*{\xleftleftarrows}[2][]{%
  \ext@arrow 3095\leftleftarrowsfill@{#1}{#2}%
}
\makeatother
\nc{\xrra}{\xrightrightarrows}
\nc{\xlla}{\xrightrightarrows}

\usepackage{enumitem}
\usepackage[colorlinks=true,hyperindex, linkcolor=magenta, pagebackref=false, citecolor=cyan,pdfpagelabels]{hyperref}
\usepackage[capitalize]{cleveref}
\setlist[enumerate]{itemsep=2pt,parsep=2pt,before={\parskip=2pt}}
\usepackage{xurl}

\usepackage{makecell}
\usepackage{multirow}

\usepackage{longtable}
\usepackage{booktabs}
\usepackage{url}
\usepackage{mathrsfs}
\usepackage{boxedminipage}
\usepackage{anyfontsize}
\usepackage{listings}
\usepackage{xcolor}
\newcommand{\RNum}[1]{\uppercase\expandafter{\romannumeral #1\relax}}
\usepackage{bbold}
\usepackage{wrapfig}
\usepackage{multicol}
\usepackage{extarrows}
\usepackage{titletoc}
\usepackage{dynkin-diagrams}
\usepackage{mathtools}
\usepackage{enumitem}
\usepackage[usestackEOL]{stackengine}

\usepackage{pdfpages}
\makeatletter
\DeclareFontFamily{OMX}{MnSymbolE}{}
\DeclareSymbolFont{MnLargeSymbols}{OMX}{MnSymbolE}{m}{n}
\SetSymbolFont{MnLargeSymbols}{bold}{OMX}{MnSymbolE}{b}{n}
\DeclareFontShape{OMX}{MnSymbolE}{m}{n}{
    <-6>  MnSymbolE5
   <6-7>  MnSymbolE6
   <7-8>  MnSymbolE7
   <8-9>  MnSymbolE8
   <9-10> MnSymbolE9
  <10-12> MnSymbolE10
  <12->   MnSymbolE12
}{}
\DeclareFontShape{OMX}{MnSymbolE}{b}{n}{
    <-6>  MnSymbolE-Bold5
   <6-7>  MnSymbolE-Bold6
   <7-8>  MnSymbolE-Bold7
   <8-9>  MnSymbolE-Bold8
   <9-10> MnSymbolE-Bold9
  <10-12> MnSymbolE-Bold10
  <12->   MnSymbolE-Bold12
}{}

\let\llangle\@undefined
\let\rrangle\@undefined
\DeclareMathDelimiter{\llangle}{\mathopen}%
                     {MnLargeSymbols}{'164}{MnLargeSymbols}{'164}
\DeclareMathDelimiter{\rrangle}{\mathclose}%
                     {MnLargeSymbols}{'171}{MnLargeSymbols}{'171}
\makeatother

\numberwithin{equation}{section}
\newtheorem{theorem}{Theorem}
\numberwithin{theorem}{section}
\newtheorem{thm}[theorem]{Theorem}

\newtheorem{prop/defi}[theorem]{Proposition/Definition}
\newtheorem{lem/defi}[theorem]{Lemma/Definition}
\newtheorem{thm/defi}[theorem]{Theorem/Definition}
\newtheorem{defi/prop}[theorem]{Definition/Proposition}
\theoremstyle{definition}
\newtheorem{defi}[theorem]{Definition}

\newtheorem{rem}[theorem]{Remark}

\newtheorem{defi/rem}[theorem]{Definition/Remark}

\makeatletter
\patchcmd{\@tocline}
  {\hfil}
  {\leaders\hbox{\,.\,}\hfil}
  {}{}
\makeatother

\makeatletter
    \def\cleardoublepage{\clearpage%
        \if@twoside
            \ifodd\c@page\else
                \vspace*{\fill}
                \hfill
                \begin{center}
                This page is intentionally left blank.
                \end{center}
                \vspace{\fill}
                \thispagestyle{empty}
                \newpage
                \if@twocolumn\hbox{}\newpage\fi
            \fi
        \fi
    }
\makeatother

\usepackage[new]{old-arrows}
\usepackage{verbatim}
\usepackage{bookmark}
\usepackage{scalerel,stackengine}
\stackMath
\newcommand\rwh[1]{%
\savestack{\tmpbox}{\stretchto{%
  \scaleto{%
    \scalerel*[\widthof{\ensuremath{#1}}]{\kern-.6pt\bigwedge\kern-.6pt}%
    {\rule[-\textheight/2]{1ex}{\textheight}}
  }{\textheight}%
}{0.5ex}}%
\stackon[1pt]{#1}{\tmpbox}%
}
\providecommand*{\rightrightarrowsfill@}{%
  \arrowfill@\relrelbar\relrelbar\rightrightarrows
}
\providecommand*{\leftleftarrowsfill@}{%
  \arrowfill@\leftleftarrows\relrelbar\relrelbar
}
\providecommand*{\xrightrightarrows}[2][]{%
  \ext@arrow 0359\rightrightarrowsfill@{#1}{#2}%
}
\providecommand*{\xleftleftarrows}[2][]{%
  \ext@arrow 3095\leftleftarrowsfill@{#1}{#2}%
}
\makeatother

\begin{document}

\title[Characterizations of sharp solute-solvent interfaces]{Characterizations of sharp solute-solvent interfaces\\in hydrophobic environments via cylindrical coordinates}
\author{Hao XIAO}
\address[Previous]{School of Mathematical Sciences, Soochow University, Suzhou, Jiangsu 215006, China}
\email{hxiao@stu.suda.edu.cn}
\address[Current]{Mathematisches Institut, Universit\"at Bonn, Endenicher Allee 60, D-53115, Bonn, Germany}
\email{xiao@uni-bonn.de}
\begin{abstract}
This paper characterizes sharp solute-solvent interfaces in hydrophobic environments, and there are three major ingredients. The first is the variational implicit solvent model (VISM) which establishes the free energy functional of arbitrary solvation states. The minimization of this functional yields a PDE which characterizes both the stable and saddle solute-solvent interfaces. The second is a solute system consisting of two Gay-Berne ellipsoidal hydrophobic molecules. The cylindrical coordinates reduce the corresponding PDE to an ODE which is more traceable, increasing the computability of the solute-solvent interfaces under solvation effects. The third, which is the innovative part of this paper, transforms the computation of sharp solute-solvent interfaces into an ODE boundary value problem. We introduce an effective method for numerically computing saddle interfaces. This remains a hard problem in previous research. Besides, we address some qualitative results and study how the interface varies with respect to the distance between the two Gay-Berne ellipsoidal hydrophobic molecules.

\smallskip\noindent\textbf{Keywords}:
Solute-solvent interface;
Gay-Berne ellipsoidal hydrophobic molecule;
Minimization of free energy functional;
Numerical ODE boundary value problem.
\end{abstract}
\maketitle
\tableofcontents

\st{Introduction and background}

\noindent
The study of hydrophobic phenomena is important in biology, chemistry, and biochemistry. For a long time, the interactions between solutes and solvents in hydrophobic environments have been considered crucial to understand many important issues \cite{Bal}. For example, amphiphilic molecules self-assembling into micelles and membranes \cite{HGGP}, evaporation in capillaries \cite{LuzL,LeuL,LLB,LumL}, protein folding \cite{Fer,BOW,DSK,BGOW}, and gas solubility are all significant examples of hydrophobic hydration and interaction. As the surfaces of many biological molecules are extended nonpolar areas, it is interesting to study the aggregation and exclusion of water between these hydrophobic surfaces.

In theoretical research of hydrophobic phenomena, simplified but effective computational models are frequently used. Through theoretical analysis and molecular dynamics (MD) simulations, people have achieved numerous substantial explanations of hydrophobicity.

It is well known that the hydration of small hydrophobic solutes differs from that of large ones. Small hydrophobic solutes (\eg argon and methane) can integrate into water's hydrogen bond network without disrupting it \cite{Bal,PC,Ch1,St,PRB1,PRB2}. In contrast, larger hydrophobic solutes (\eg small molecules like neopentane \cite{HMB1}) reorganize the hydrogen bond network, causing OH groups to orient towards the molecular surfaces, and lead to the formation of dangling hydrogen bonds \cite{LMR}. Simulations have shown that, on hydrophobic protein surfaces, some hydrophobic residues do not disrupt adjacent water hydrogen bonds, while others result in ``large'' or ``small'' heterogeneous structures on protein surfaces \cite{CR}.

F.~Stillinger conjectured that large strongly hydrophobic particles would induce a liquid-vapor interface \cite{St}. Experimental evidence has shown that a vapor layer of molecular size exists around large paraffin-like molecules \cite{JJRBPKB}. This vapor layer was also predicted by the Lum-Chandler-Weeks theory \cite{LCW,HC}. By computer simulations, A.~Wallqvist and B.~Berne demonstrated that, when two strongly hydrophobic plates (Gay-Berne ellipsoids) allow multiple water layers to fill in the gap between them, some water layers will be excluded from the gap \cite{WB1,WB2}. X.~Huang, C.~Margulis, and B.~Berne, via molecular dynamics simulations, indicated that the number of excluded water layers depends on the size of the plates \cite{HMB2}.

However, J.~Dzubiella, J.~Swanson, and J.~McCammon proposed a model of sharp interfaces, namely the variational implicit solvent approach \cite{DSM1,DSM2}. This model does not consider the possible vapor layers (between solutes and solvents) but treats the solute-solvent transition as an ``abrupt'' change. This makes it possible to apply tools like differential geometry, variational methods, and partial differential equations to study hydrophobic phenomena, enabling preciser characterizations of solute-solvent interfaces to some extent. Based on their model, this paper explores the interface between solvent and solute systems composed of dual Gay-Berne ellipsoidal hydrophobic molecules.

\st{Variational implicit solvent method}

\noindent
In this section, we establish the variational implicit solvent model proposed by Dzubiella-Swanson-McCammon \cite{DSM1,DSM2}. This is an important and fruitful mathematical tool for analyzing solvation effects, and it admits a wide range of applications. To this end, we also refer to the review on this model by Cheng-Dzubiella-McCammon-Li in \cite{CDML}.

\sst{Geometric setup}\label{sst:geo}

Consider a solvation system consisting of two parts:
\begin{\enu}[label=\textup{(\roman*)}]\setcounter{enumi}{0}
\item A kind of solute of arbitrary shape and composition;
\item A kind of solvent encompassing the solute.
\end{\enu}
Let $W$ denote the domain of the entire solvation system, encompassing both the solute and solvent. Let $V$ denote the solute region (or cavity region) which excludes the solvent, and the solute-solvent interface is defined as the boundary of the solute region $V$, denoted $\Gm:=\pa V$.

In this model, the solute-solvent interface is sharp, \ie the solute-solvent transition occurs as a discrete ``jump''. We also assume that the interface $\Gm$ may contain multiple connected components (each of which is closed and continuous).

\begin{defi}
The {\bf volume exclusion function} for the cavity region $V$ is
\[v(\vec{r})
=\lt\{\begin{array}{ll}
0 & \vec{r}\in V \\
1 & \tet{else}
\end{array}\rt..\]
\end{defi}
\begin{rem}
In mathematics, the function $v(\vec{r})$ is commonly called the characteristic function of the solvent region. In physics, $v(\vec{r})$ serves as the phase field of the solvation system which describes the solute-solvent distribution and characterizes their interface.
\end{rem}
The volume $\mm{Vol}[V]$ of the solute region $V$ and the area $\mm{Area}[\Gm]$ of the interface $\Gm$ are expressed as functionals of the volume exclusion function $v(\vec{r})$, \ie
\begin{align*}
\mm{Vol}[V]&=\int_Vd^3r
=\int_W[1-v(\vec{r})]d^3r \\
\tet{and}\ \mm{Area}[\Gm]
&=\int_\Gm dS
=\int_W|\nab v(\vec{r})|d^3r
\end{align*}
where $\nab\equiv\nab_{\vec{r}}$ is the usual gradient operator with respect to the position vector $\vec{r}$ and $|\nab v(\vec{r})|$ is the $\de$-function concentrated on the boundary $\Gm=\pa V$ of the cavity region. Consequently, the expression $dS\equiv|\nab v(\vec{r})|d^3r$ can be understood as the infinitesimal surface element. Note that, in the framework of the variational implicit solvent model, both the volume exclusion function $v(\vec{r})$ and its boundary $\Gm$ can serve as the variable for describing the free energy of the solvation system.

We further assume that both the position vector of each solute atom $\vec{r}_i$ and the solute configurations are fixed. Hence, for the solvent, the solute can be treated as an external potential with no degrees of freedom. In this continuum solvent model, the solvent density distribution is given by $\rho(\vec{r})=\rho_0v(\vec{r})$ where $\rho_0$ is the volume density of the solvent.

\sst{Free energy functional}
\label{sst:functional}

Based on Subsection~\ref{sst:geo}, we apply the Gibbs free energy model proposed in \cite{DSM1,DSM2} for a solvation system:
\begin{defi}
Given a solvation system determined by its cavity region $V$ with boundary $\Gm=\pa V$ (\ie the solute-solvent interface), its {\bf Gibbs free energy} is the functional of the volume exclusion function $v(\vec{r})$ (or equivalently its boundary $\Gm$), \ie
\begin{align}\label{eq:Gibbs}
G[v]
&=G_{\mm{vol}}[v]+G_{\mm{sur}}[v]
+G_{\mm{vdW}}[v]+G_{\mm{ele}}[v] \nonumber \\
&=P\int_W[1-v(\vec{r})]d^3r
+\int_W\gm(\vec{r},\Gm)|\nab v(\vec{r})|d^3r
+\rho_0\int_W U(\vec{r})v(\vec{r})d^3r
+G_{\mm{ele}}[v].
\end{align}
\end{defi}

Now we explain (\ref{eq:Gibbs}) term by term.

The first term
\begin{equation}\label{eq:GibbsVol}
G_{\mm{vol}}[v]
:=P\int_W[1-v(\vec{r})]d^3r
=P\,\mm{Vol}[V]
\end{equation}
is the volume contribution proportional to $\mm{Vol}[V]$ where $P=P_l-P_v$ is the volume pressure difference between the liquid and vapor phases. So $G_{\mm{vol}}[v]$ describes the amount of energy required to form the cavity region $V$ in the solvent by overcoming the pressure difference $P$. Normally in water or other fluids nearly of multiphase coexistence like, this pressure difference $P$ is very small. For solutes of microscopic size ($\sim$nm), this contribution can often be neglected.

The second term
\begin{equation}\label{eq:GibbsSur}
G_{\mm{sur}}[v]
:=\int_W\gm(\vec{r},\Gm)|\nab v(\vec{r})|d^3r
=\int_\Gm\gm(\vec{r},\Gm)dS
\end{equation}
describes the energy required for solvent rearrangement around the cavity $V$ (particularly near the solute-solvent interface $\Gm$). It is expressed via the function $\gm(\vec{r},\Gm)$ and the interface $\Gm$. Note that this energy penalty is regarded as the primary factor causing hydrophobic phenomena \cite{Ch2}. It depends on the properties of the solvent and also relies on the specific topological structure of the interface $\Gamma$, varying locally in space \cite{TZ}.

The closed-form expression of the function $\gm(\vec{r},\Gm)$ is unknown. The following first-order curvature correction approximation based on scaled particle theory \cite{Re} was proposed in \cite{DSM1,DSM2}:
\begin{equation}\label{eq:curvature}
\gm(\vec{r},\Gm)=\gm_0[1-2\tau H(\vec{r})]
\end{equation}
where $\gm_0$ is the constant surface tension of the liquid-vapor interface of the solvent and $\tau$ is a positive constant commonly referred to as the Tolman length \cite{To}. We have the local mean curvature
\[
H(\vec{r})=\fr{1}{2}
[\kap_1(\vec{r})+\kap_2(\vec{r})]
\]
where $\kap_1(\vec{r})$ and $\kap_2(\vec{r})$ are the local principal curvatures of the interface $\Gm$.

Molecular dynamics (MD) simulations have shown that the surface tension $\gm_0$ is the asymptotic value of the spin energy per unit surface area of a rigid spherical cavity of large radius limit in water \cite{LCW,HGC}. In this solvation system, the Tolman length is estimated to be of molecular size in the range $0.7$--$0.9$\AA. Since its precise value is unknown, the Tolman length can serve as the sole fitting parameter in the variational continuum solvent model. The mean curvature $H$ is only defined on the solute-solvent interface $\Gm$. By convention, we let the curvature be positive for convex surfaces (\eg a spherical cavity) 
and negative for concave surfaces (\eg a spherical droplet).

For the geometric contribution of the free energy functional (\ref{eq:Gibbs}), we provide an intuitive description based on 
the Hadwiger theorem in differential geometry \cite{Ha}. We recall that the Gaussian curvature of a general smooth surface is $K(\vec{r})=\kap_1(\vec{r})\kap_2(\vec{r})$.
\begin{thm}[Hadwiger]
\label{thm:Hadwiger}
Let $C$ denote the set of all convex bodies in $\RR^3$, and let $M$ be the collection of all finite unions of convex bodies in $C$. Suppose a functional $F:M\ra\RR$ satisfies:
\begin{\enu}[label=\textup{(\roman*)}]\setcounter{enumi}{0}
\item Invariance under rigid transformations;
\item Additivity: $F(U \cup V) = F(U) + F(V) - F(U \cap V)$ for all $U,V\in M$;
\item Continuity: If $U_i,U\in M$ and $U_i\ra U$, then $F(U_i)\ra F(U)$.
\end{\enu}
Then we have
\[
F(U)
=a\,\mm{Vol}[U]
+b\,\mm{Area}[\pa U]
+c\int_{\pa U}H(\vec{r})dS
+d\int_{\pa U}K(\vec{r})dS
\]
for all $U\in M$.
\end{thm}
\begin{rem}
In \thrm{thm:Hadwiger}, a functional $F:M\ra\RR$ satisfying the conditions (i)--(iii) is often called a valuation function of the elements of $M$ or simply a valuation on $M$.
\end{rem}

By physical intuition, we expect that the free energy $G[v]$, as a functional of $v(\vec{r})$ or the interface $\Gm$, satisfies the conditions (i)--(iii) in \thrm{thm:Hadwiger}. Thus, according to \thrm{thm:Hadwiger}, the geometric contribution of the free energy $G[v]$, as a valuation of the closed surface 
$\Gm$, should include all terms of $G_{\mm{vol}}+G_{\mm{sur}}$ (\ie the volume, surface area, and surface integral of mean curvature) as well as the surface integral of the Gaussian curvature $K(\vec{r})$ over $\Gm$. 

We review the Gauss-Bonnet theorem \cite[pp. 267]{doC}:
\begin{thm}[Gauss-Bonnet]
\label{thm:Gauss-Bonnet}
Let $\Gm=\bigsqcup\Gm_i$ be a disjoint union of smooth closed surfaces, then the surface integral of the Gaussian curvature $K(\vec{r})$ over $\Gm$ is given by
\[
\int_\Gm K(\vec{r})dS
=2\pi\sum_i\chi_e(\Gm_i)
\]
where $\chi_e(\Gm_i)$ is the Euler characteristic of the surface $\Gm_i$.
\end{thm}

Since the Gaussian curvature $K(\vec{r})$ is an intrinsic geometric property of the surface $\Gm$, its surface integral contributes an additive constant to the free energy. So it does not affect the minimization process of the energy. However, it should be noted that, if the topology of the interface $\Gamma$ changes, the integral $\int_\Gm K(\vec{r})dS$ would capture it. So we assume that the minimization of the free energy $G[v]$ is carried out for a fixed topology on the interface.

The third term  
\begin{equation}\label{eq:GibbsvdW}
G_{\mm{vdW}}[v]
:=\rho_0\int_WU(\vec{r})v(\vec{r})d^3r
=\rho_0\int_{W\sm V} U(\vec{r})d^3r
\end{equation}
is the total non-electrostatic solute-solvent interaction energy of van der Waals type for a given solvent density distribution $\rho_0v(\vec{r})$. The potential energy
\begin{equation}\label{eq:potential}
U(\vec{r})
=\sum_{i=1}^NU_i(|\vec{r}-\vec{r}_i|)
\end{equation}
is the sum of $U_i$'s, which describes the interaction of the $i$-th solute atom centered at $\vec{r}_i$ (and all $N$ solute atoms) with the surrounding solvent. Each $U_i$ contains the short-range repulsion and the long-range dispersion between the solute atom at $\vec{r}_i$ and the solvent molecule at $\vec{r}$. The classical opinion often takes $U_i$ as the isotropic Lennard-Jones (LJ) potential
\begin{equation}\label{eq:LJ}
U_{\mm{LJ}}(r)
=4\ep\lt[\lt(\fr{\sig}{r}\rt)^{12}
-\lt(\fr{\sig}{r}\rt)^6\rt]
\end{equation}
where $\ep$ is the energy scale, $\sig$ is the length scale, and $r$ is the central distance (\ie the distance between centers). In Subsection~\ref{sst:single}, we will get some corrections to (\ref{eq:LJ}) for Gay-Berne ellipsoidal molecules.

The final term $G_{\mm{ele}}[v]$ is the electrostatic energy arising from charges possibly produced by solute atoms and ions in the solvent. In this paper, we focus on nonpolar solutes. So we omit this term in subsequent discussions. Based on this assumption and (\ref{eq:Gibbs})--(\ref{eq:potential}), we arrive at the final form of the nonpolar free energy functional:
\begin{equation}\label{eq:GibbsFinal}
G[v]=P\,\mm{Vol}[V]
+\int_\Gm\gm_0[1-2\tau H(\vec{r})]dS
+\rho_0\sum_{i=1}^N\int_{W\sm V}
U_i(|\vec{r}-\vec{r}_i|)d^3r
\end{equation}
where every interaction potential $U_i$ has the form as (\ref{eq:LJ}). Additionally, we let
\begin{align*}
G_{\text{area}}[v]
&=\gm_0\int_{\Gm}dS
=\gm_0\,\text{Area}[\Gamma] \\
\tet{and}\ G_{\text{mean}}[v]
&=-2\tau\gm_0\int_{\Gm}H(\vec{r})dS.
\end{align*}

\sst{Minimization of free energy functional}

Let $v_{\min}(\vec{r})$ or its boundary $ \Gamma_{\min}$ be the volume exclusion function that minimizes (\ref{eq:GibbsFinal}), then the Gibbs free energy of the solvation system is given by $G[v_{\min}]$. The solvation free energy $\Delta G$ is the reversible effect of the solute solvation given by
\[
\Delta G=G[v_{\min}]-G_0
\]
where $G_0$ is a constant reference energy determined by pure solvent or unsolvated solid. Up to an additive constant, the solvent-mediated potential energy of the average force along a given reaction coordinate $x$ (\eg the distance between the mass centers of two solutes) is given by
\[
w(x)=G[v_{\min}]
\]
where $v_{\min}(\vec{r})$ must be computed separately for each $x$.

A necessary condition for the solute-solvent interface $\Gm$ to minimize the energy is that the first variation of the free energy functional (\ref{eq:GibbsFinal}) vanishes at the corresponding volume exclusion function $v$, \ie at all points on the boundary $\Gm$ we have
\begin{equation}\label{eq:partialG}
\fr{\de G[v]}{\de v}=0.
\end{equation}
This kind of energy change can be regarded as a distribution on the interface $\Gm$, and it is given by the following equation \cite{DSM1,DSM2,OH,OS}:
\begin{equation}
\label{eq:partialGDistribution}
\fr{\de G[v]}{\de v}
=P+2\gm_0[H(\vec{r})-\tau K(\vec{r})]
-\rho_0U(\vec{r}).
\end{equation}
The PDE derived from (\ref{eq:partialG}) and (\ref{eq:partialGDistribution}) that determines the optimal exclusion function $v_{\min}(\vec{r})$ (or the optimal solute-solvent interface $\Gm_{\min}$) is expressed through pressure, curvature, short-range repulsion, and long-range dispersion. These quantities all have the dimension of energy density. It can interpret the mechanical balance between the forces contributed by each term on the right-hand side of (\ref{eq:partialGDistribution}) per unit surface area on the surface $\Gm$. In fact, solving this equation in three dimensions is very difficult and complicated.

\begin{rem}
It is worth noting that the minimization of the free energy functional $G[v]$ is fundamentally different from the minimal surface problem presented in \cite{Sp}. The latter focuses on minimizing the surface area of a surface with a given closed curve as its boundary, whereas the former considers not only geometric part $G_{\mm{vol}}+G_{\mm{sur}}$ but also other functional minimization conditions such as $G_{\mm{vdW}}[v]$ and $G_{\mm{ele}}[v]$ in polar environments.
\end{rem}

\begin{rem}
According to L.~Landau, every physical process corresponds to the variation of the integral of a certain Lagrange function of the system \cite[pp. 1--2]{LaL}. Noting that the formation of the solute-solvent interface under solvation is a physical process, the free energy functional $G[v]$ can thus be understood as the integral of the Lagrange function of the solvation system. Based on Hamilton's principle of least action \cite[pp. 2]{LaL}, the variation of this integral leads to the Euler-Lagrange equation \cite[pp. 3]{LaL} given by
\begin{equation}\label{eq:EL}
P+2\gm_0[H(\vec{r})
-\tau K(\vec{r})]-\rho_0U(\vec{r})=0.
\end{equation}
\end{rem}

\st{Solvation system of dual Gay-Berne ellipsoidal hydrophobic solute molecules}

\noindent
In this section, we first establish the solvation system of dual Gay-Berne ellipsoidal hydrophobic solute molecules based on \cite{HMB2,HZB}. Then we simplify (\ref{eq:EL}) using cylindrical coordinates to improve the computability of the solute-solvent interface.

\sst{Lennard-Jones potential of a single Gay-Berne ellipsoidal hydrophobic molecule}
\label{sst:single}

We start by describing the Lennard-Jones potential induced by a single Gay-Berne ellipsoidal hydrophobic molecule. Let the ellipsoidal hydrophobic molecule be perpendicular to the $z$-axis, and let the $z$-axis pass through the center of the ellipsoid. Suppose $P$ is an arbitrary point in space, $\vec{r}$ denotes the position vector of $P$, $\oli{r}$ denotes the distance between $P$ and the center $O$ of the ellipsoid, and $r$ denotes the distance between $P$ and the $z$-axis. (The abuse of the symbol $r$ shall not cause any confusion in the context and does not conflict with the infinitesimal volume element $d^3r$.) Let $\theta$ be the angle between $\overrightarrow{OP}$ and the positive direction of the $z$-axis. The size of the ellipsoidal molecule is illustrated in the figure below.
\begin{figure}[ht!]
\centering
\includegraphics[scale=0.2]{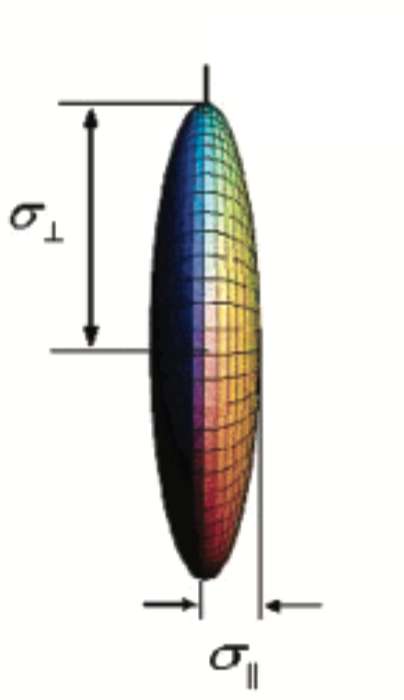}
\caption{Single Gay-Berne ellipsoidal hydrophobic solute molecule \cite{HZB}.}
\end{figure}

\noindent The Lennard-Jones potential induced by this molecule in space is given by
\begin{equation}\label{eq:ellipsoidLJ}
U(\vec{r})
=U_{\mm{LJ}}(\oli{r},\ta)
=4\ep\lt[
\lt(\fr{\sig_0}{\oli{r}-\sig(\chi;\ta)+\sig_0}\rt)^{12}-
\lt(\fr{\sig_0}{\oli{r}-\sig(\chi;\ta)+\sig_0}\rt)^6
\rt]
\end{equation}
where $\sig(\chi;\ta):=\sig_{\perp}(1-\chi\cos^2\ta)^{-1/2}$ and $\chi:=({\sig_{\parallel}}^2-{\sig_{\perp}}^2)/{\sig_{\parallel}}^2$.

\sst{Lennard-Jones potential of dual Gay-Berne ellipsoidal hydrophobic molecules}
\label{sst:double}

For two identical Gay-Berne ellipsoidal hydrophobic molecules (as depicted in Subsection \ref{sst:single}), let both ellipsoids be perpendicular to the $z$-axis with the $z$-axis passing through their centers. Assume that the two ellipsoids are located on opposite sides of the origin with their centers at a fixed (positive) distance $d$ from the origin along the $z$-axis. Then $D:=2d$ is the distance between the two ellipsoids. Label the ellipsoid on the negative $z$-axis as Ellipsoid I and the one on the positive $z$-axis as Ellipsoid II, and let their centers be denoted by $O_1$ and $O_2$ respectively. Let $P$ be an arbitrary point in space, and let $\vec{r}$ be its position vector. Denote $\oli{r}_1$ as the distance from $P$ to $O_1$ and $\oli{r}_2$ as the distance from $P$ to $O_2$. Define $\theta_1 = \angle PO_1O_2$ and $\theta_2 = \angle PO_2O_1$. The size of the two Gay-Berne ellipsoidal molecules is illustrated in the figure below.
\begin{figure}[ht!]
\centering
\includegraphics[scale=0.35]{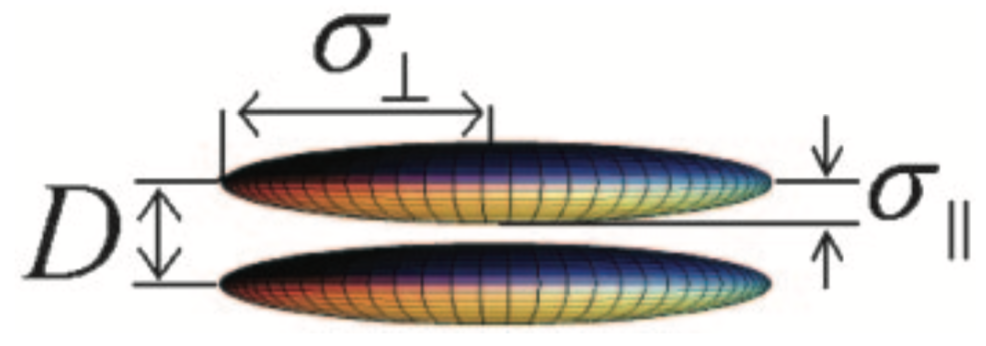}
\caption{Dual Gay-Berne ellipsoidal hydrophobic solute molecules \cite{HMB2}.}
\end{figure}

\noindent
Let $u$ be the $z$-coordinate of $P$. The Lennard-Jones potential induced by this solvation system is
\[
U(\vec{r})=U_1(\vec{r})+U_2(\vec{r})
\]
where $U_1(\vec{r})=U_{\mm{LJ}}(\oli{r}_1,\ta_1)$, $U_2(\vec{r})=U_{\mm{LJ}}(\oli{r}_2,\ta_2)$,
\begin{align*}
\oli{r}_1&=\sq{r^2+(d+u)^2},
\ \ta_1=\arctan\fr{r}{d+u}, \\
\tet{and}
\ \oli{r}_2&=\sq{r^2+(d-u)^2},
\ \ta_2=\arctan\fr{r}{d-u}.
\end{align*}

\sst{Cylindrical parameterization, curvature, and free energy minimization}
\label{sst:sym}

For the solvation model established in Subsection~\ref{sst:double}, we further assume that the solute-solvent interface formed during solvation is a surface of revolution about the $z$-axis and is symmetric with respect to the $xOy$-plane. The interface $\Gm$ can be parameterized in cylindrical coordinates as
\begin{equation}\label{eq:para}
\Gm(\vphi,t):\lt\{
\begin{array}{l}
x = r(t)\cos\vphi \\
y = r(t)\sin\vphi \\
z = z(t)
\end{array}\rt.
\end{equation}
where $\vphi$ and $t$ are parameters that determine the two principal directions and the normal direction of the surface $\Gm=\Gm(\vphi, t)$.
By \cite[pp. 163--165]{doC}, the two principal curvatures, mean curvature, and Gaussian curvature at a point $\vec{r}$ on the surface $\Gm$ are respectively given by
\[
\kap_1(\vec{r})
=\fr{\ddot{z}\dot{r}-\dot{z}\ddot{r}}
{(\dot{r}^2+\dot{z}^2)^{3/2}},
\ \kap_2(\vec{r})
=\fr{\dot{z}}{r\sq{\dot{r}^2+\dot{z}^2}},
\ H(\vec{r})=
\fr{\kap_1(\vec{r})+\kap_2(\vec{r})}{2},
\ \tet{and}\ K(\vec{r})
=\kap_1(\vec{r})\kap_2(\vec{r}).
\]
where $\dot{r} = dr/dt$, $\ddot{r} = d^2r/dt^2$, $\dot{z} = dz/dt$, and $\ddot{z} = d^2z/dt^2$.

During the minimization process, we neglect $G_{\mathrm{vol}}[v]$ (see Subsection~\ref{sst:functional}). Then (\ref{eq:EL}) under the parameterization (\ref{eq:para}) is simplified as
\begin{equation}\label{eq:var}
2H(\vec{r})-2\tau K(\vec{r})
=\fr{\rho_0}{\gm_0}U(\vec{r}),
\ \vec{r}\in\Gm
\end{equation}
where the expressions of the curvatures and potential energy are all known.

\st{Computation of sharp solute-solvent interfaces}

\noindent
This section is the innovative part of this paper. We first describe the methods for computing solute-solvent interfaces in different topological states and derive some qualitative conclusions, then we compute various numerical results of the interface. We will investigate the interface in two distinct topological states:
\begin{\enu}[label=\textup{(\roman*)}]\setcounter{enumi}{0}
\item
A single connected component, \ie a smooth closed surface contains both solute molecules;
\item
The other one of two connected components, \ie two smooth closed surfaces each of which contains exactly one solute molecule.
\end{\enu}
We assume that all connected components in the sense of topology are diffeomorphic to $S^2$.

\sst{Computation method for connected interfaces}\label{sst:connected}

We first consider the portion of the interface between two Gay-Berne ellipsoidal solute molecules. (The computation method for the other portions of the interface will be given in the next subsection.) In this case, we set $t \equiv z$ in (\ref{eq:para}) (so that the normal direction of the interface locally points towards the solvent region $W \setminus V$), and the cylindrical coordinate parameterization of $\Gm$ can be expressed as
\[
\Gm(\vphi,z):\lt\{
\begin{array}{l}
x=r(z)\cos\vphi \\
y=r(z)\sin\vphi \\
z=z(z)\equiv z
\end{array}\rt..\]
The principal curvatures are then
\[
\kap_1=-\fr{\ddot{r}}{(1+\dot{r}^2)^{3/2}}
\ \tet{and}\ \kap_2
=\fr{1}{r\sq{1+\dot{r}^2}}.
\]
Now (\ref{eq:var}) becomes
\[
\lt[\fr{2\tau}{r(1+\dot{r}^2)^2}
-\fr{1}{(1+\dot{r}^2)^{3/2}}\rt]\ddot{r}
+\fr{1}{r\sq{1+\dot{r}^2}}
=\fr{\rho_0}{\gm_0}U(z,r(z)).
\]
which is reorganized as
\begin{equation}
\label{eq:connected}
\ddot{r}
=\fr{r(1+\dot{r}^2)^2}
{2\tau-r\sq{1+\dot{r}^2}}
\lt(\fr{\rho_0}{\gm_0}
U(z,r(z))-\fr{1}{r\sq{1+\dot{r}^2}}\rt).
\end{equation}
This is a second-order nonlinear ordinary differential equation. We will numerically solve it using the bvp4c function in MATLAB. Besides, we conduct a ``reasonable abuse'' of notation by adjusting the variable of $U(\vec{r})$ depending on $t$, which shall not cause ambiguity in the context.

For the minimal solution of (\ref{eq:connected}), the $z$-domain is set to be $[-d,d]$. The boundary condition is
\[\lt\{\begin{aligned}
\dot{r}(-d)&=0,\\
\dot{r}(d)&=0.
\end{aligned}\rt.\]
When computing this portion of the surface, we always put the initial condition as
\[
r\equiv r_0,\ z\in[-d,d]
\]
where $r_0$ is a positive constant slightly greater than $\sig_{\perp}$.

For the saddle solution of (\ref{eq:connected}), inspired by \cite{DSWZ} and noting the symmetric assumption on the interface proposed in Subsection~\ref{sst:sym} as well as the constant fitting capability of the bvp4c function, we consider the portion of the interface where its $z$-coordinate falls in the interval $[0,d]$. We use the minimum value $r_{\min}$ of $r(z)$ where $-d\le z\le d$ as boundary condition on the left, and we take $d$ as an undetermined constant. After performing the transformation $s = z/d$, (\ref{eq:connected}) becomes
\begin{equation}\label{eq:saddle}
\fr{d^2r}{ds^2}
=\fr{s^2r(1+(\dot{r}/s)^2)^2}
{2\tau-r\sq{1+(\dot{r}/s)^2}}
\lt(\fr{\rho_0}{\gm_0}U(d\cdot s,r(d\cdot s))
-\fr{1}{r\sq{1+(\dot{r}/s)^2}}\rt).
\end{equation}
Here, the $s$-domain is $[0, 1]$, and the boundary condition is
\[\lt\{\begin{aligned}
r(0)&=r_{\min} \\
\dot{r}(1)&=0
\end{aligned}\rt..\]
When computing this portion of the interface, we always use the initial condition
\[
r\equiv r_0,\ s\in[0,1]
\]
where $r_0$ is a positive constant slightly greater than $\sigma_{\perp}$.

\sst{Computation method for disconnected interfaces}

In this subsection, we consider the computation of disconnected interfaces. (This method is also applicable when interface is connected and for regions where the $z$-coordinate of the interface lies outside of $[-d, d]$). By letting $t\equiv r$ in (\ref{eq:para}) where the local normal direction of the interface points towards the solvent region $W\sm V$, the cylindrical parameterization of $\Gm$ is
\begin{equation}\label{eq:disconnected}
\Gm(\vphi, r):\left\{
\begin{array}{l}
x=r(r)\cos\vphi\equiv r\cos\vphi \\
y=r(r)\sin\vphi\equiv r\sin\vphi \\
z=z(r)
\end{array}\right..
\end{equation}
In this case, the principal curvatures are
\[
\kap_1=\fr{\ddot{z}}{(1+\dot{z}^2)^{3/2}}
\ \tet{and}\ \kap_2
=\fr{\dot{z}}{r\sq{1+\dot{z}^2}}.
\]
Then (\ref{eq:var}) becomes
\[
\lt[\fr{1}{(1+\dot{z}^2)^{3/2}}
-\fr{2\tau\dot{z}}
{r(1+\dot{z}^2)^2}\rt]\ddot{z}
= \fr{\rho_0}{\gm_0}U(r,z(r))
-\fr{\dot{z}}{r\sq{1+\dot{z}^2}}
\]
which is reorganized as
\begin{equation}\label{eq:down}
\ddot{z}
=\fr{r(1+\dot{z}^2)^2}{r\sq{1+\dot{z}^2}
-2\tau\dot{z}}
\left(\fr{\rho_0}{\gm_0}U(r,z(r))
-\fr{\dot{z}}{r\sq{1+\dot{z}^2}}\right).
\end{equation}

\noindent
Note that, under the parameterization (\ref{eq:disconnected}) due to the normal direction of the surface, we can only compute the portion of the interface that are concave downward on the $rOz$-coordinate plane. To compute the portion of the interface that are concave upward on the $rOz$-coordinate plane, we need to simultaneously change the signs of both principal curvatures of the surface. In this case, the principal curvatures are
\[
\kap_1
=-\fr{\ddot{z}}{(1+\dot{z}^2)^{3/2}}
\ \tet{and}\ \kap_2
=-\fr{\dot{z}}{r\sq{1+\dot{z}^2}}.
\]
And (\ref{eq:var}) becomes
\begin{equation}\label{eq:up}
\ddot{z}
=-\fr{r(1+\dot{z}^2)^2}
{r\sq{1+\dot{z}^2}+2\tau\dot{z}}
\lt(\fr{\rho_0}{\gm_0}U(r,z(r))
+\fr{\dot{z}}{r\sq{1+\dot{z}^2}}\rt).
\end{equation}
In both situations, the boundary condition is
\[\left\{\begin{aligned}
\dot{z}(0)&=0\\
z(d)&=r_{\max}
\end{aligned}\right.\]
where $r_{\max}$ is determined by solving (\ref{eq:connected}) or (\ref{eq:saddle}), \ie the maximum value of $r(z)$ solving (\ref{eq:connected}) for $z\in[-d,d]$ or that of $r(s)$ solving (\ref{eq:saddle}) for $s\in[0,1]$.

We set the initial condition of (\ref{eq:down}) as
\[
z=d-\sig_\perp
\lt(\fr{\sig_\parallel-r}
{\sig_\parallel}\rt)^{1/4},
\ r\in[0,1].
\]
The initial condition of (\ref{eq:up}) is
\[
z=d+\sig_\perp\lt(\fr{\sig_\parallel - r}{\sig_\parallel}\rt)^{1/4},\ r\in[0,1].
\]

\sst{Connectedness criterion for solute-solvent interfaces}

In this subsection, we present a sufficient condition that determines the connectedness of the solute-solvent interface. Theoretically, this confirms that no solvent exists between two Gay-Berne ellipsoidal hydrophobic molecules. Conversely, it also demonstrates that our simplified model coincides with physical observations to some extent. This result is inspired by techniques handling boundary value conditions in \cite{SK}.

Suppose a disconnected interface exists, then the interface admits the parameterization (\ref{eq:disconnected}) and (\ref{eq:down}) has a solution. Moreover, we assume that $z$ as a function of $r$ is smooth. Note that $z=z(r)$ satisfies $z(0)=p$ and $\dot{z}(0)=0$ at $r=0$ where $p$ is a constant to be determined.

By Taylor expansion, when $r\ra0^+$ we have:
\begin{align*}
z(r)&=z(0)+\dot{z}(0)r
+\frac{1}{2}\ddot{z}(0)r^2
+\cdots\ssim p
+\frac{1}{2}\ddot{z}(0)r^2 \\
\tet{and}\ \dot{z}(r)&=\dot{z}(0)
+\ddot{z}(0)r
+\cdots\ssim\ddot{z}(0)r.
\end{align*}
By smoothness of $z=z(r)$, we deduce
\begin{align*}
\ddot{z}(0)
&=\lim_{r\to0^+}\fr{r(1+\dot{z}^2)^2}
{r\sqrt{1+\dot{z}^2}-2\tau\dot{z}}
\lt(\fr{\rho_0}{\gm_0}U(r,z(r))-\fr{\dot{z}}{r\sqrt{1+\dot{z}^2}}\rt)\\
&=\lim_{r\to0^+}\fr{(1+\dot{z}^2)^2}
{\sqrt{1+\dot{z}^2}
-2\tau\fr{\dot{z}(r)-\dot{z}(0)}{r-0}}
\lt(\fr{\rho_0}{\gm_0}U(r,z(r))
-\fr{\fr{\dot{z}(r)-\dot{z}(0)}{r-0}}
{\sqrt{1+\dot{z}^2}}\rt)\\
&=\fr{1}{1-2\tau\ddot{z}(0)}
\lt(\fr{\rho_0}{\gm_0}
U(0,p)-\ddot{z}(0)\rt).
\end{align*}
Thus, we obtain a quadratic equation for $\ddot{z}(0)$:
\begin{equation}\label{eq:quadratic}
2\tau\left(\ddot{z}(0)\right)^2
-2\ddot{z}(0)
+\frac{\rho_0}{\gm_0}U(p)=0.
\end{equation}
It has a solution only if
\[
U(p)\le\fr{\gm_0}{2\tau\rho_0}.
\]
Regarding the symmetric assumption on the interface proposed in Subsection~\ref{sst:sym} and the properties of the Lennard-Jones potential, we conclude
\begin{thm}[Connectedness criterion for solute-solvent interfaces]
\label{thm:connected}
If $U({\bold 0})>\fr{\gm_0}{2\tau\rho_0}$, the solute-solvent interface has only one connected component which is diffeomorphic to $S^2$.
\end{thm}

\begin{rem}
When the distance between the two ellipsoidal hydrophobic molecules is not too small, \thrm{thm:connected} implies that, if the hydrophobicity of the ellipsoidal molecules is sufficiently strong (\ie the Lennard-Jones potential induced by the solute system in the region
\[
\{r=0,-d\le z\le d\}
\]
is positive), the solute-solvent interface contains only one connected component. This is consistent with physical intuition.
\end{rem}

\begin{rem}
Numerical computations suggest $\ddot{z}(0)=0$. So we just put $\ddot{z}(0)=0$. Combining this with (\ref{eq:quadratic}), we find that the solutions of $U(0,0,z)=0$ for $z\in[-d,d]$ characterize the intersection points of the disconnected interface with the $z$-axis. If $U(0,0,z)=0$ holds precisely when $z=0$, numerical computations indicate that the solvation system admits a saddle interface which is homeomorphic to $(S^2\sqcup S^2)/\{{\bold0}_1\sim{\bold0}_2\}$. Note that $(S^2\sqcup S^2)/\{{\bold0}_1\sim{\bold0}_2\}\not\ssim S^2$, so this saddle interface differs from the saddle one described in Subsection~\ref{sst:connected}.
\end{rem}

\sst{Numerical computation of the interface for \texorpdfstring{$D=10$}{TEXT}}

We take $D=10$ for the distance between ellipsoidal molecules and perform numerical computations for three types of solute-solvent interfaces under two topological states. The parameter values we use are shown in the table below. The parameters involving the solvent (\ie $\rho_0$, $\gm_0$, and $\tau$) are chosen to match real water.

\begin{figure}[ht!]
\centering
\begin{minipage}[c]{7.26cm}
\centering
\includegraphics[scale=0.4]{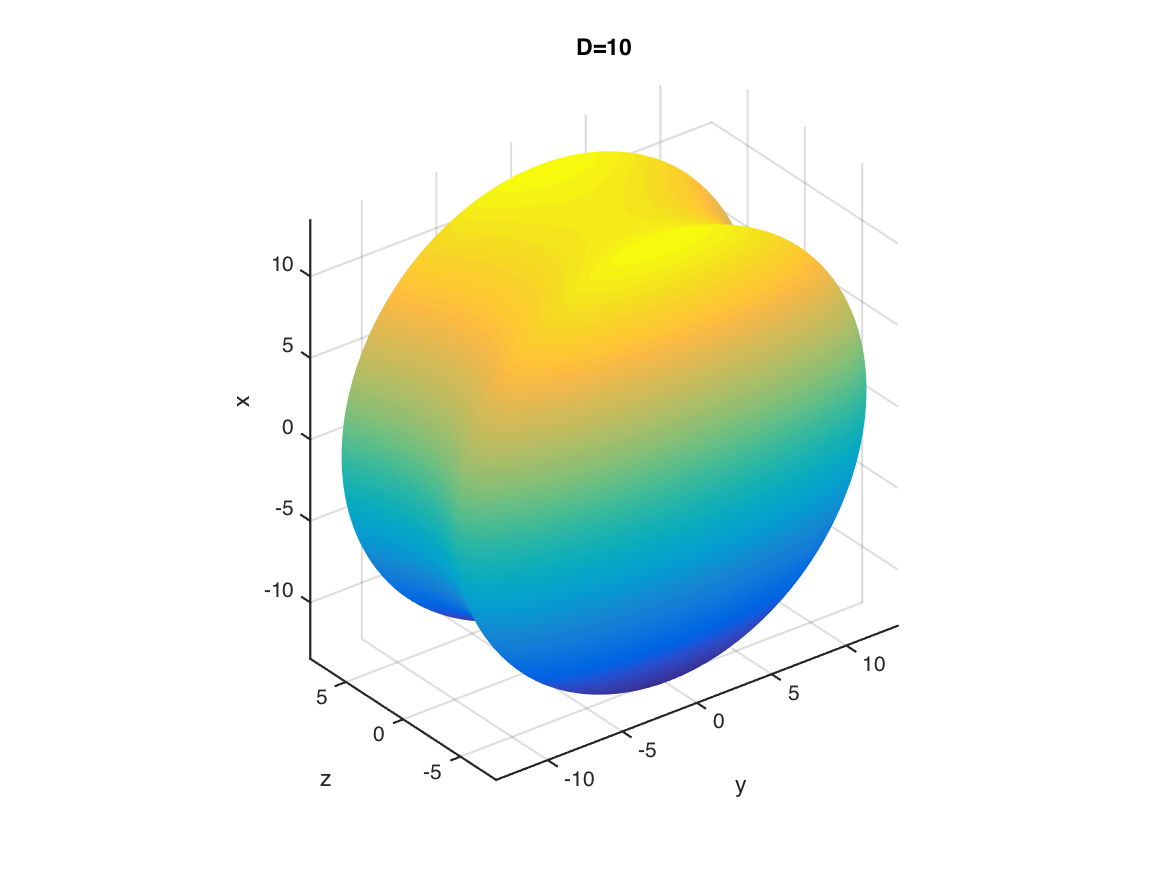}
\end{minipage}
\begin{minipage}[c]{7.26cm}
\centering
\includegraphics[scale=0.4]{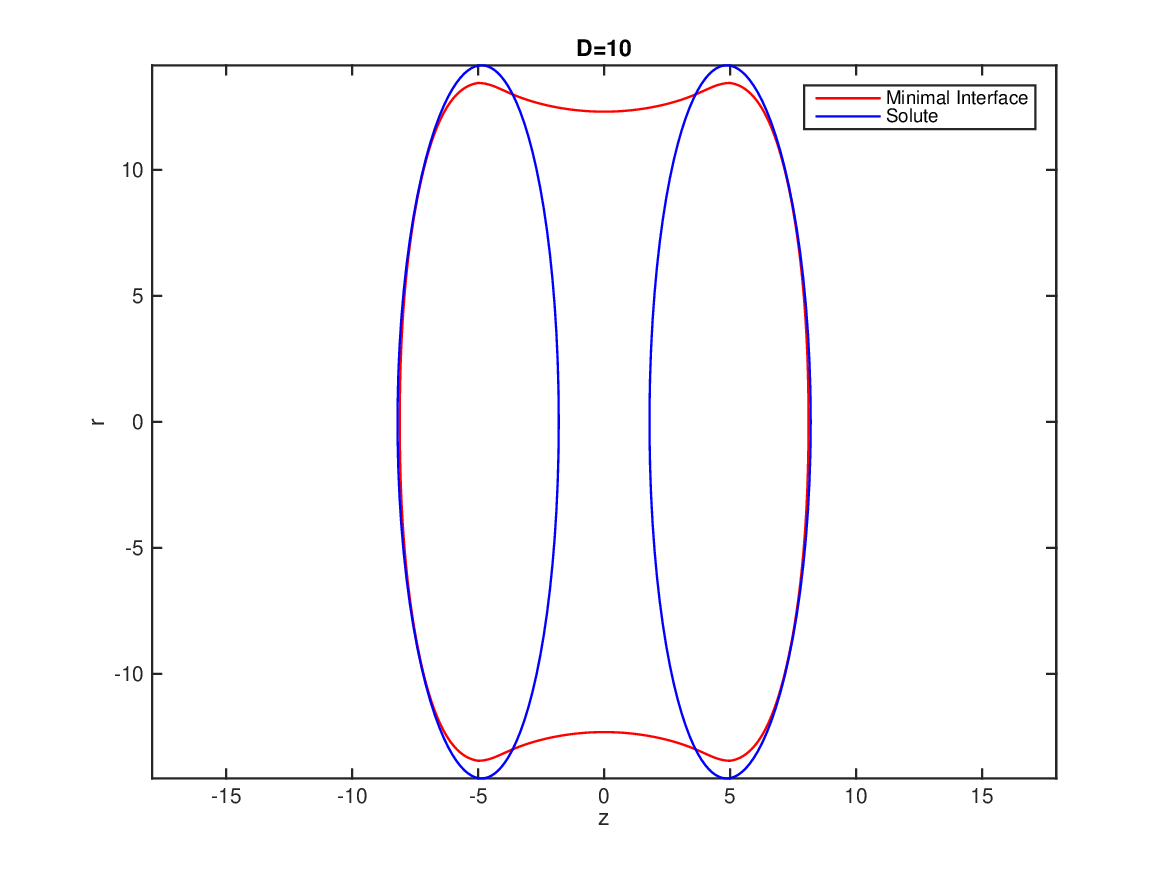}
\end{minipage}
\caption{\label{Fig:D=10_min} For $D=10$, the stable connected solute-solvent interface (left) and its cross section on the $rOz$ plane (right).}
\end{figure}

\begin{figure}[ht!]
\centering
\begin{minipage}[c]{7.26cm}
\centering
\includegraphics[scale=0.4]{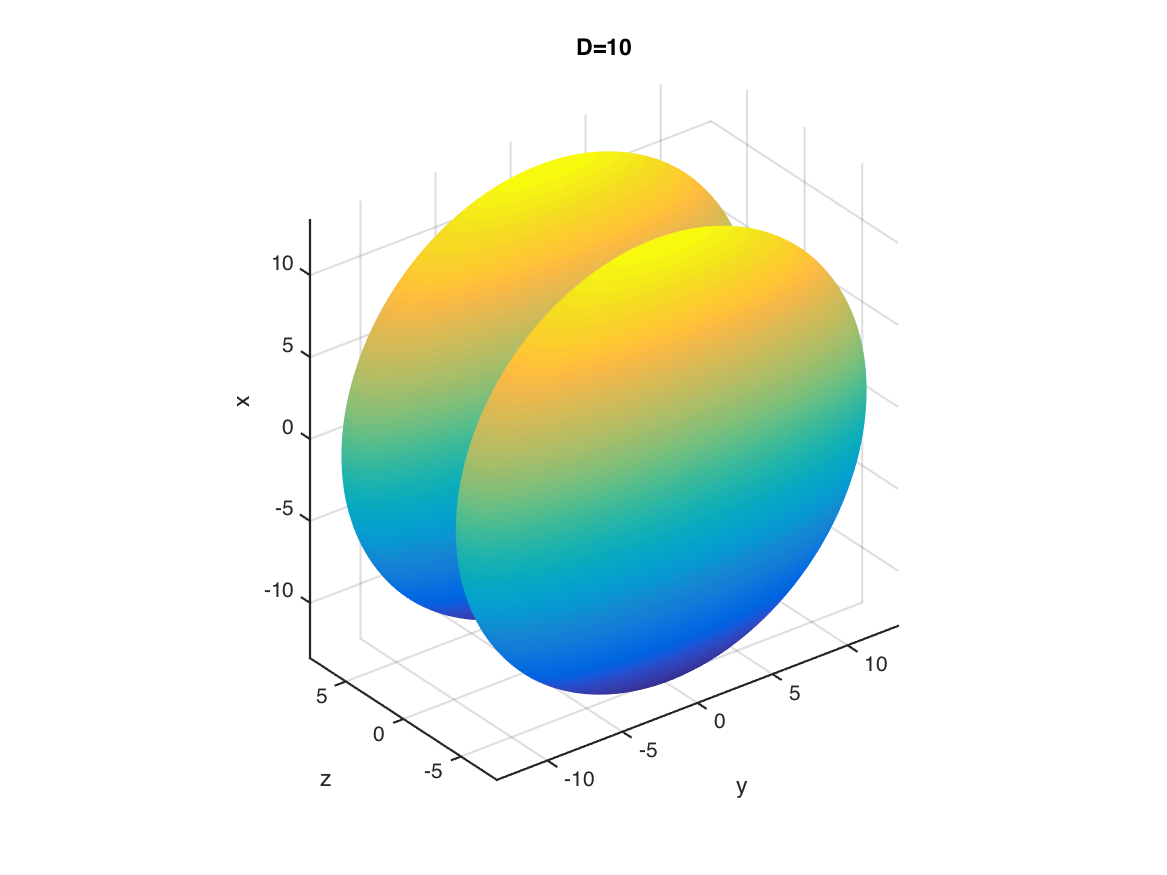}
\end{minipage}
\begin{minipage}[c]{7.26cm}
\centering
\includegraphics[scale=0.4]{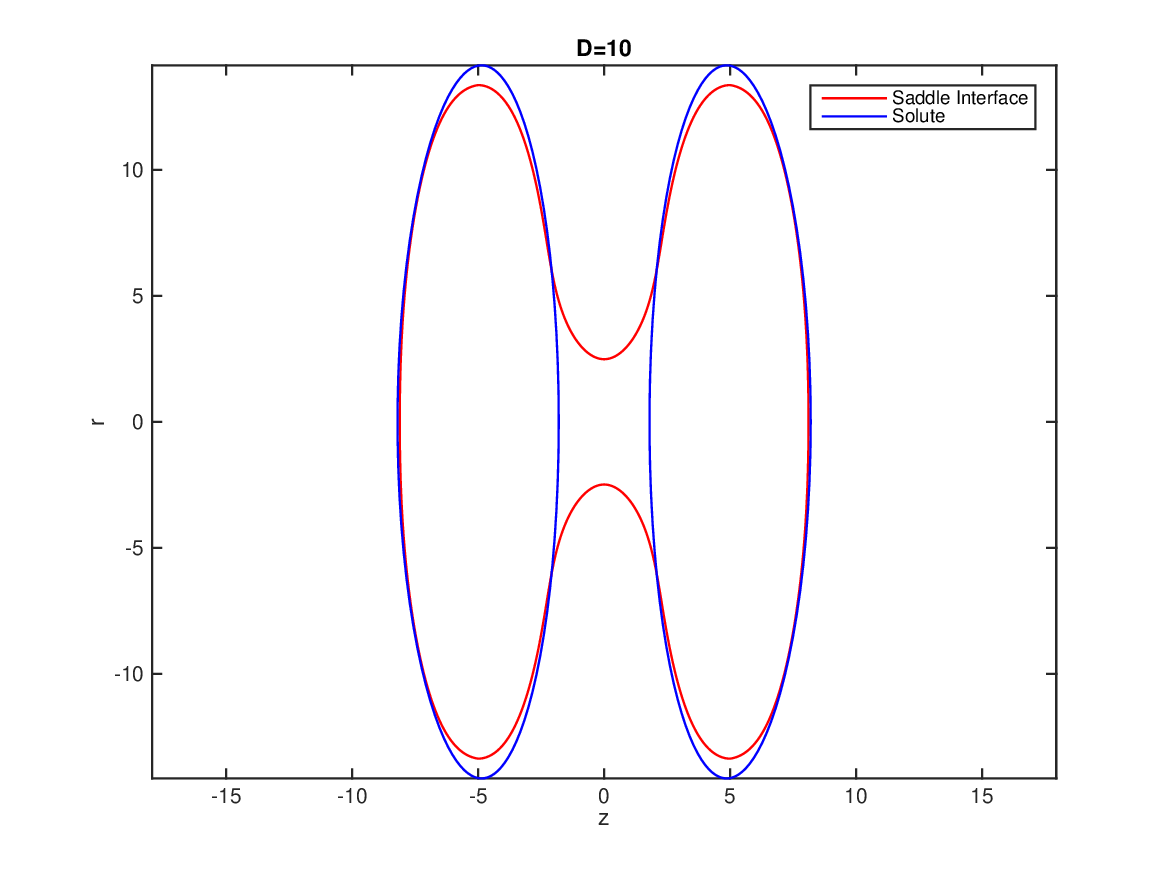}
\end{minipage}
\caption{\label{Fig:D=10_saddle} For $D=10$, the saddle connected solute-solvent interface (left) and its cross section on the $rOz$ plane (right).}
\end{figure}

\begin{figure}[ht!]
\centering
\begin{minipage}[c]{7.26cm}
\centering
\includegraphics[scale=0.4]{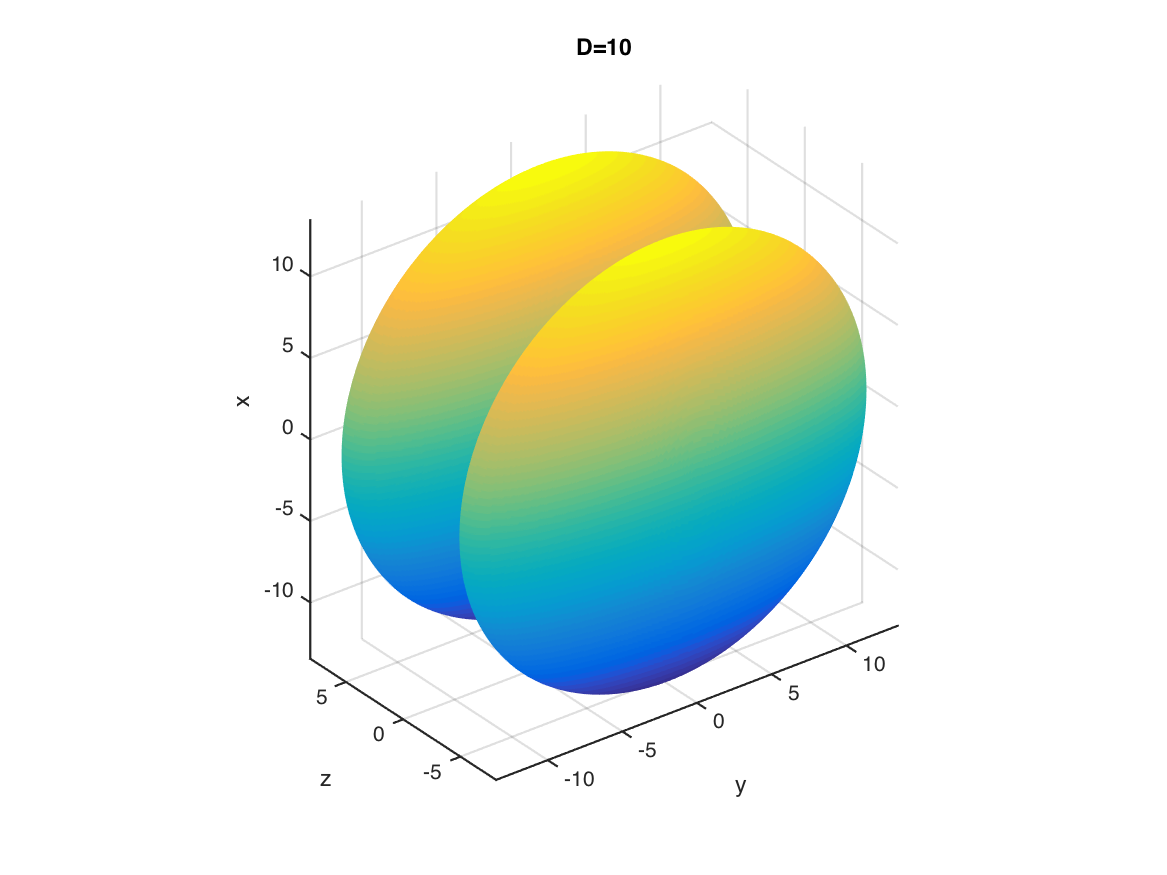}
\end{minipage}
\begin{minipage}[c]{7.26cm}
\centering
\includegraphics[scale=0.4]{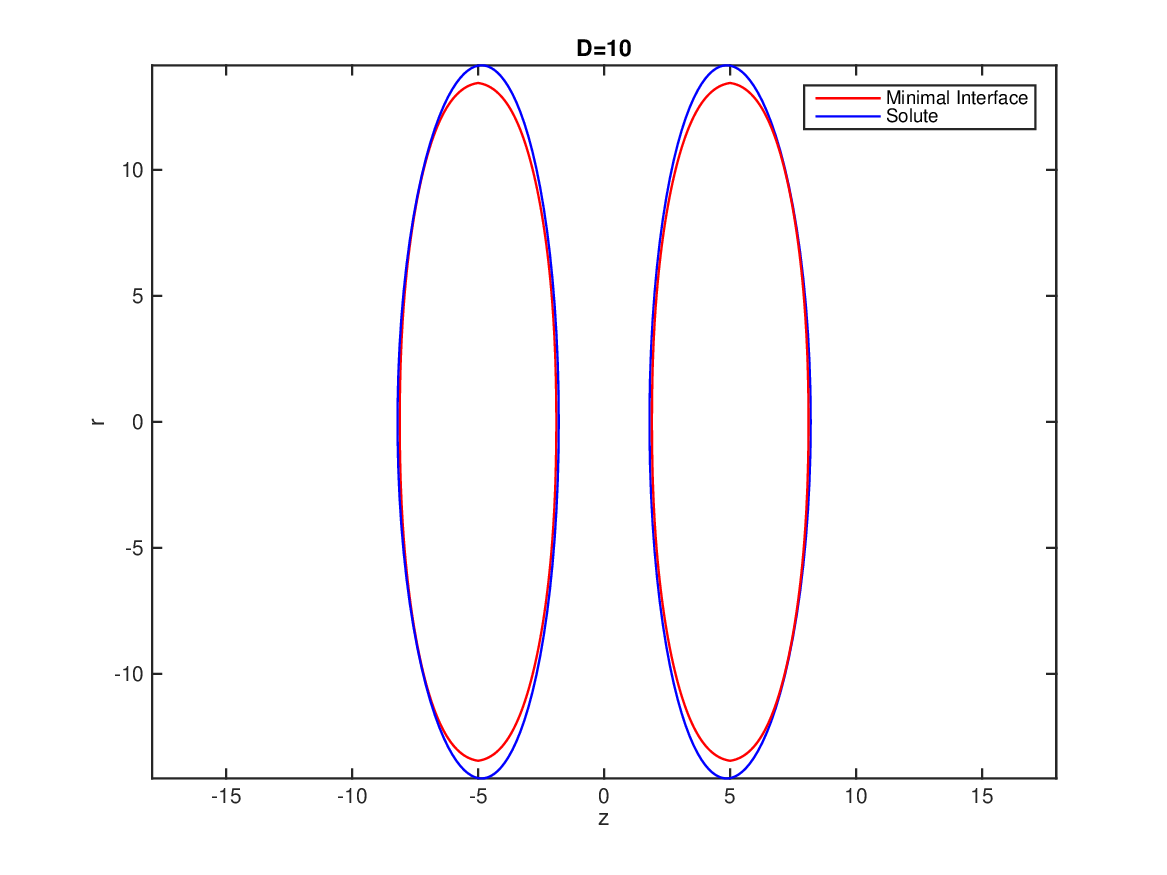}
\end{minipage}
\caption{\label{Fig:D=10_dis} For $D=10$, the stable disconnected solute-solvent interface (left) and its cross section on the $rOz$ plane (right).}
\end{figure}

Through numerical simulations, the three types of interfaces for $D=10$ correspond to Figure~\ref{Fig:D=10_min}, \ref{Fig:D=10_saddle}, and \ref{Fig:D=10_dis} respectively. (Note that, in this paper, all length units are in {\AA} and all energy units are in k$_\mm{B}$T.) We find that (\ref{eq:saddle}) can accurately compute the saddle interface (Figure~\ref{Fig:D=10_saddle}), which remains a challenging task in previous research.

\begin{table}[ht!]
\begin{center}
\begin{tabular}{c|cll}
Name & Symbol & Value & Unit \\ \hline
Volume density of solvent & $\rho_0$ & $0.033$ & \AA$^{-3}$ \\
Constant surface tension of solvent liquid-vapor interface & $\gamma_0$ & $0.175$ & k$_\mm{B}$T$\cdot$\AA$^{-2}$ \\
Tolman length & $\tau$ & $0.8$ & \AA \\
Energy scale & $\ep$ & $0.25$ & k$_\mm{B}$T \\
Equatorial radius of Gay-Berne ellipsoid & $\sigma_\perp$ & $14.1$ & \AA \\
Polar radius of Gay-Berne ellipsoid & $\sigma_\parallel$ & $3.2$ & \AA \\
Length scale of Gay-Berne ellipsoid & $\sigma_0$ & $4.6$ & \AA
\end{tabular}
\caption{\label{tab} Parameter values for computing the solvation system of dual Gay-Berne ellipsoidal hydrophobic solute molecules \cite{HMB2,CDML,HZB}.}
\end{center}
\end{table}

\sst{The interface morphology varying by the distance between ellipsoidal molecules}

Let the minimum distance from a point on the solute-solvent interface (the portion between two ellipsoidal molecules) to the $z$-axis be denoted by $r_{\min}$. By numerical computations, we obtain the graph of $r_{\min}$ varying with respect to the distance $D$ between ellipsoidal molecules, and we verify that the computation method for the (connected) saddle interface also works for the (connected) stable interface.

In numerical computations, we found that, when accurate to two decimal places for $D>8.26$, the system no longer has a connected solute-solvent interface. Similarly for $r_{\min}<1.61$, the system no longer has a saddle solute-solvent interface.

\begin{figure}[ht!]
\centering
\begin{minipage}[c]{7.26cm}
\centering
\includegraphics[scale=0.4]{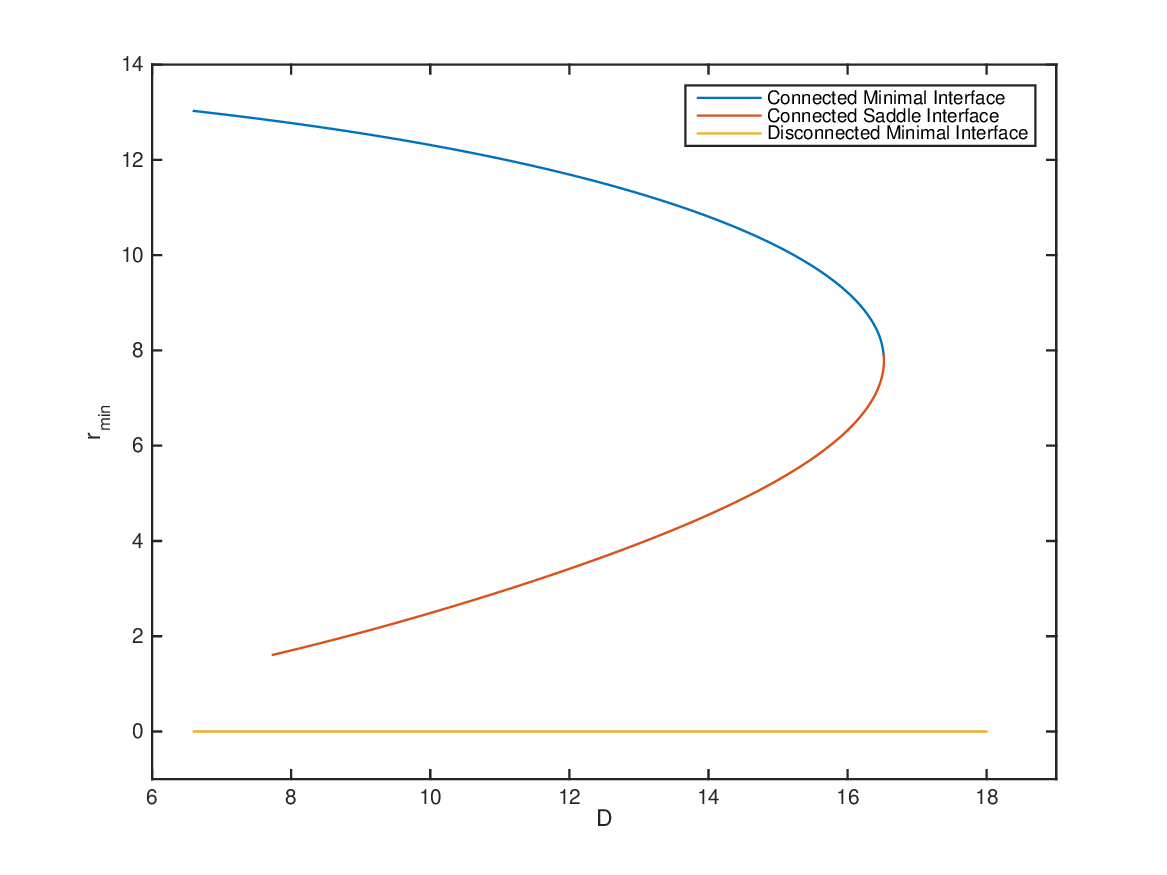}
\end{minipage}
\begin{minipage}[c]{7.26cm}
\centering
\includegraphics[scale=0.4]{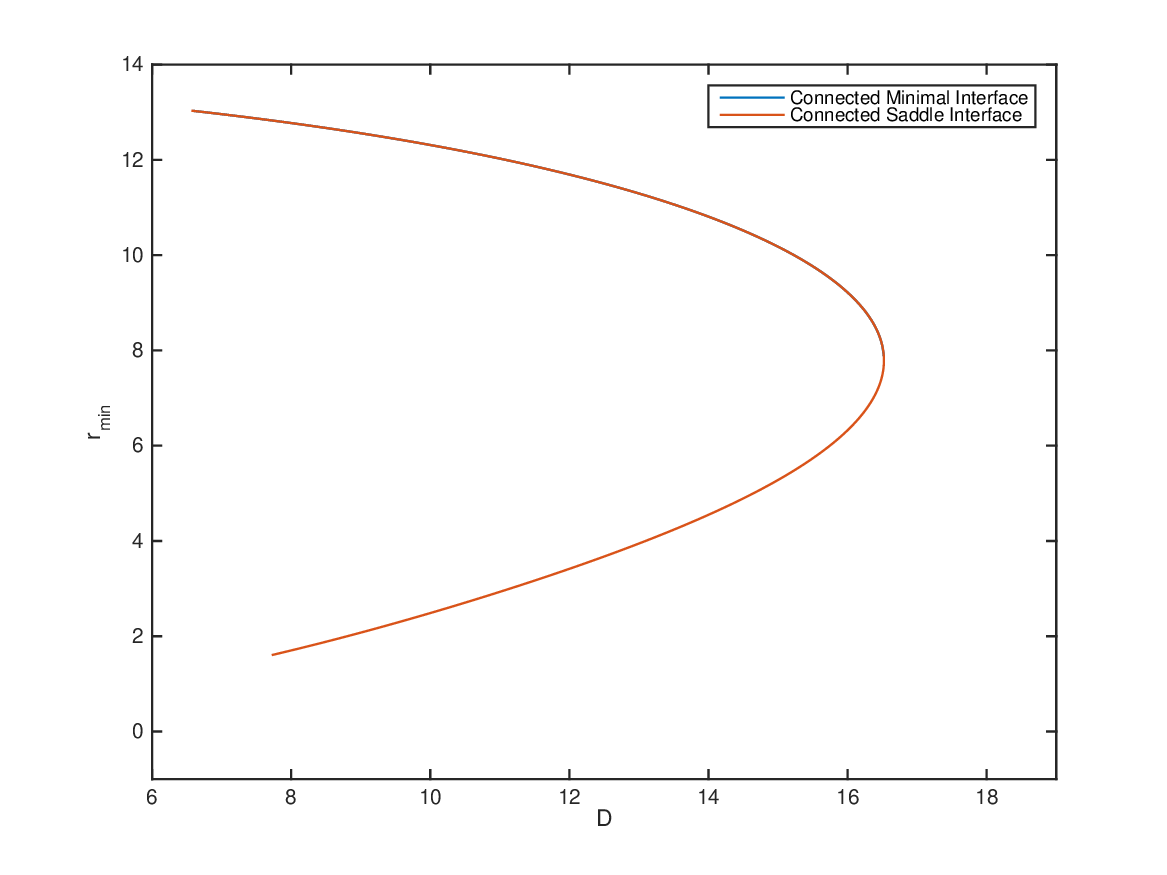}
\end{minipage}
\caption{\label{Fig:r_min}The varying trend of $r_{\min}$ with respect to $D$ (left); Simulation of the entire $r_{\min}$-$D$ curve of the saddle interface (right) where the stable one coincides with the upper half of the saddle one.}
\end{figure}

From (\ref{eq:LJ}), we observe that as a point in space gradually moves away from the solute system, the potential energy there rapidly approaches $0$\,k$_\text{B}$T. Therefore, in numerical computations, we let the region $W$ of the entire system as a cylinder with $L=50$, \ie
\[W:\lt\{
\begin{array}{l}
x=r\cos\varphi \\
y=r\sin\varphi \\
z\end{array}
\rt.,
\,0\le r\le L,
\,0\le\vphi\le2\pi,
\,-L\le z\le L.\]

\begin{figure}[ht!]
\centering
\includegraphics[scale=0.625]{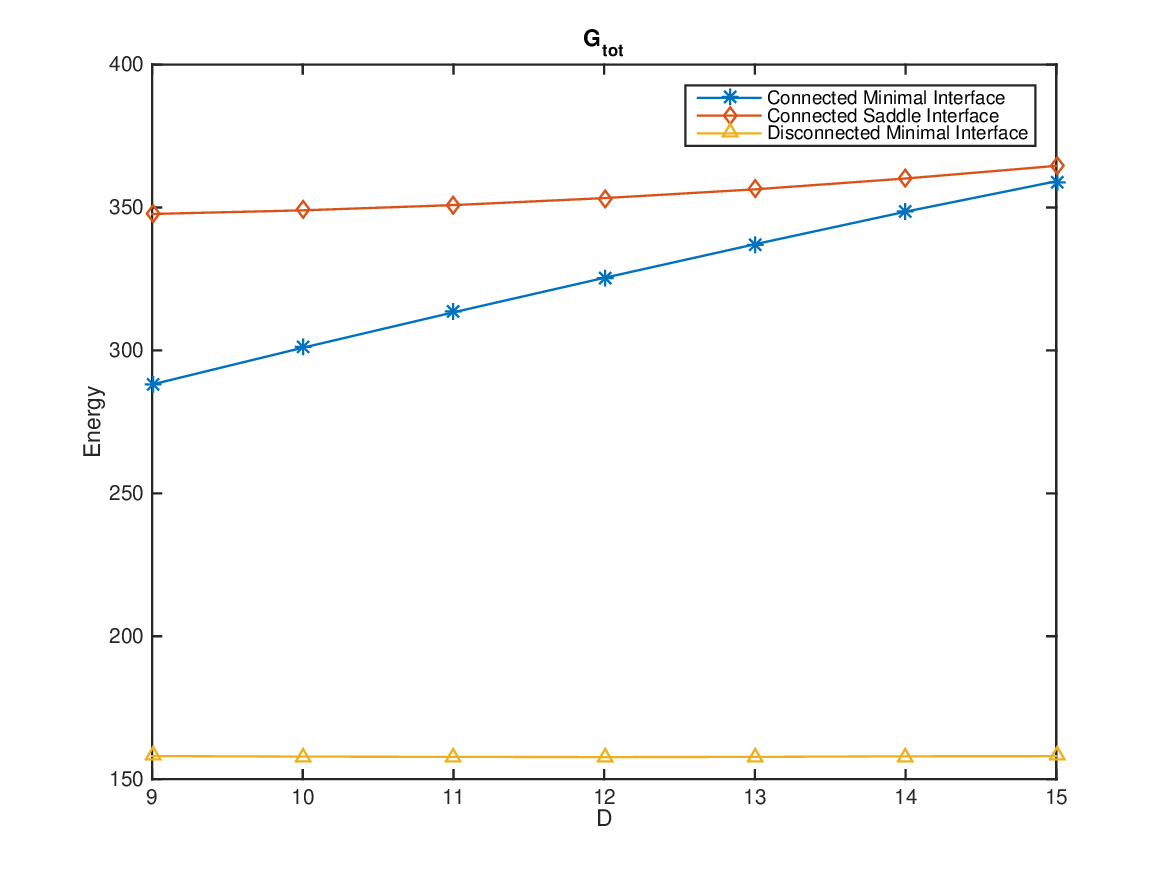}
\caption{\label{Fig:G_tot}The varying trend of $G_{\text{tot}}$ with respect to $D$.}
\end{figure}

We study the graphs of $G_{\mm{area}}[v]$, $G_{\mm{mean}}[v]$, $G_{\mm{geo}}[v]$, $G_{\mm{vdW}}[v]$, and $G_{\mm{tot}}[v]:=G[v]$ varying with respect to $D$. The numerical results are shown in the figures below. Figure~\ref{Fig:G_not_tot_2} confirms the prediction in \cite{Ch2} that $G_{\mm{sur}}[v]$ is the primary cause of hydrophobic phenomena. (Note that, in this paper, we have $G_{\mm{sur}}[v]=G_{\mm{geo}}[v]$.) Moreover, Figure \ref{Fig:G_tot} shows that for a fixed $D$ the total energy of the saddle interface is the highest, the total energy of the connected stable interface is slightly lower than that of the saddle state interface. However, the energy of the non-connected stable interface is rather low and almost does not change when $D$ varies. This is consistent with physical intuition.

\begin{figure}[ht!]
\centering
\begin{minipage}[c]{7.26cm}
\centering
\includegraphics[scale=0.4]{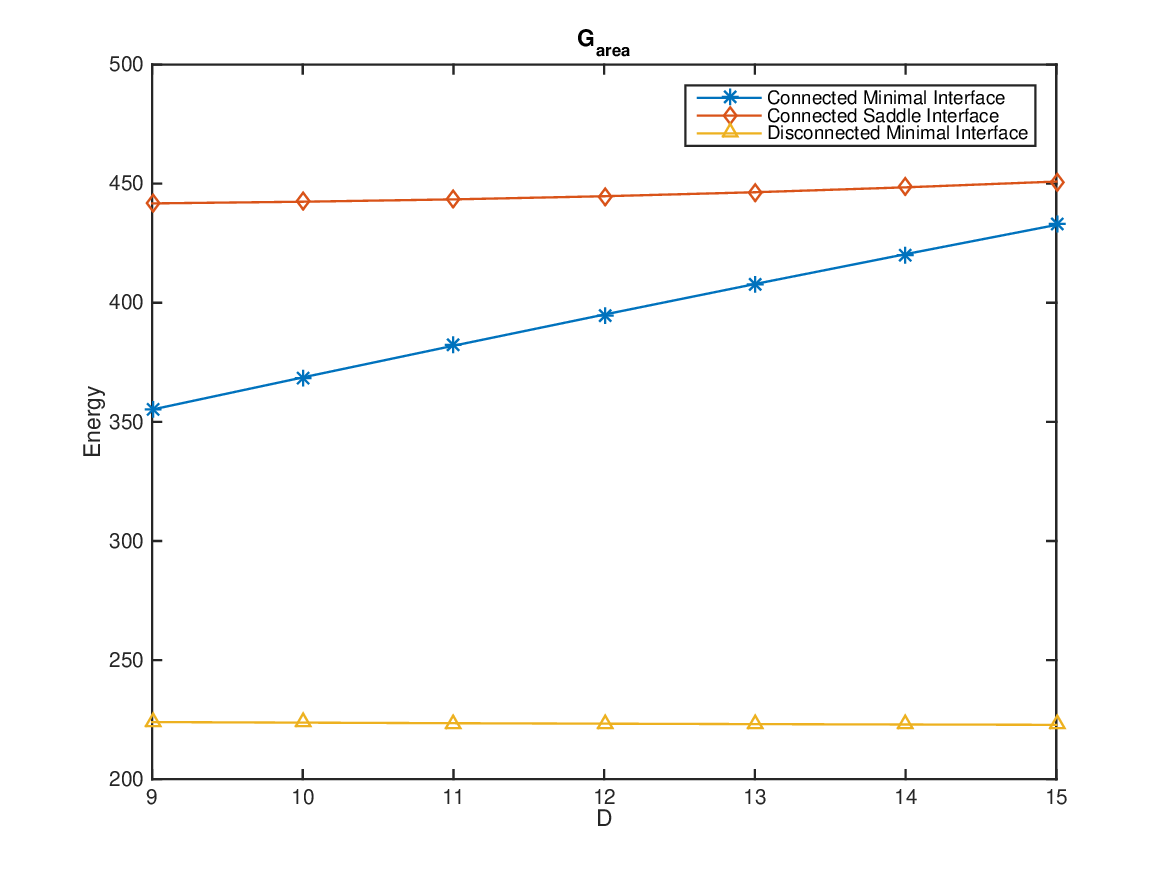}
\end{minipage}
\begin{minipage}[c]{7.26cm}
\centering
\includegraphics[scale=0.4]{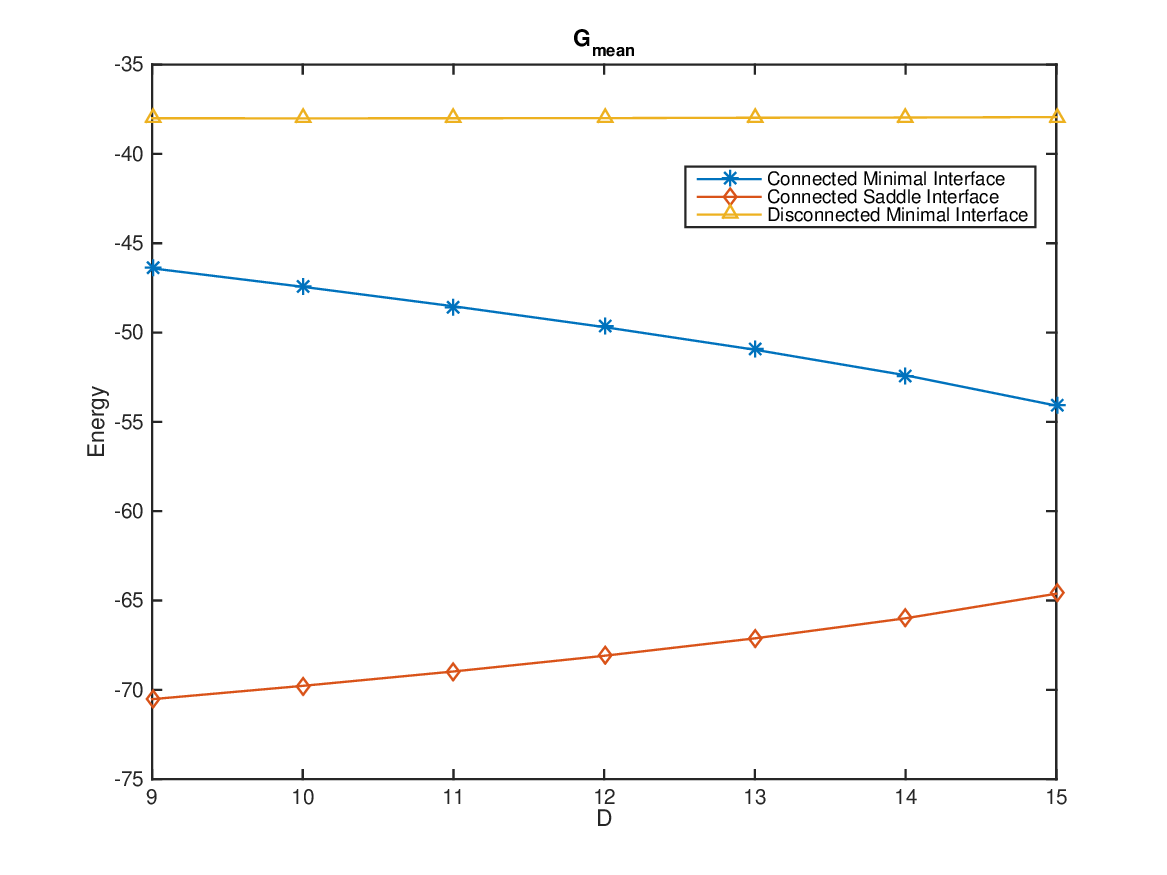}
\end{minipage}
\caption{\label{Fig:G_not_tot_1}The varying trend of $G_{\text{area}}$ with respect to $D$ (left); The varying trend of $G_{\text{mean}}$ with respect to $D$ (right).}
\end{figure}

\begin{figure}[ht!]
\centering
\begin{minipage}[c]{7.26cm}
\centering
\includegraphics[scale=0.4]{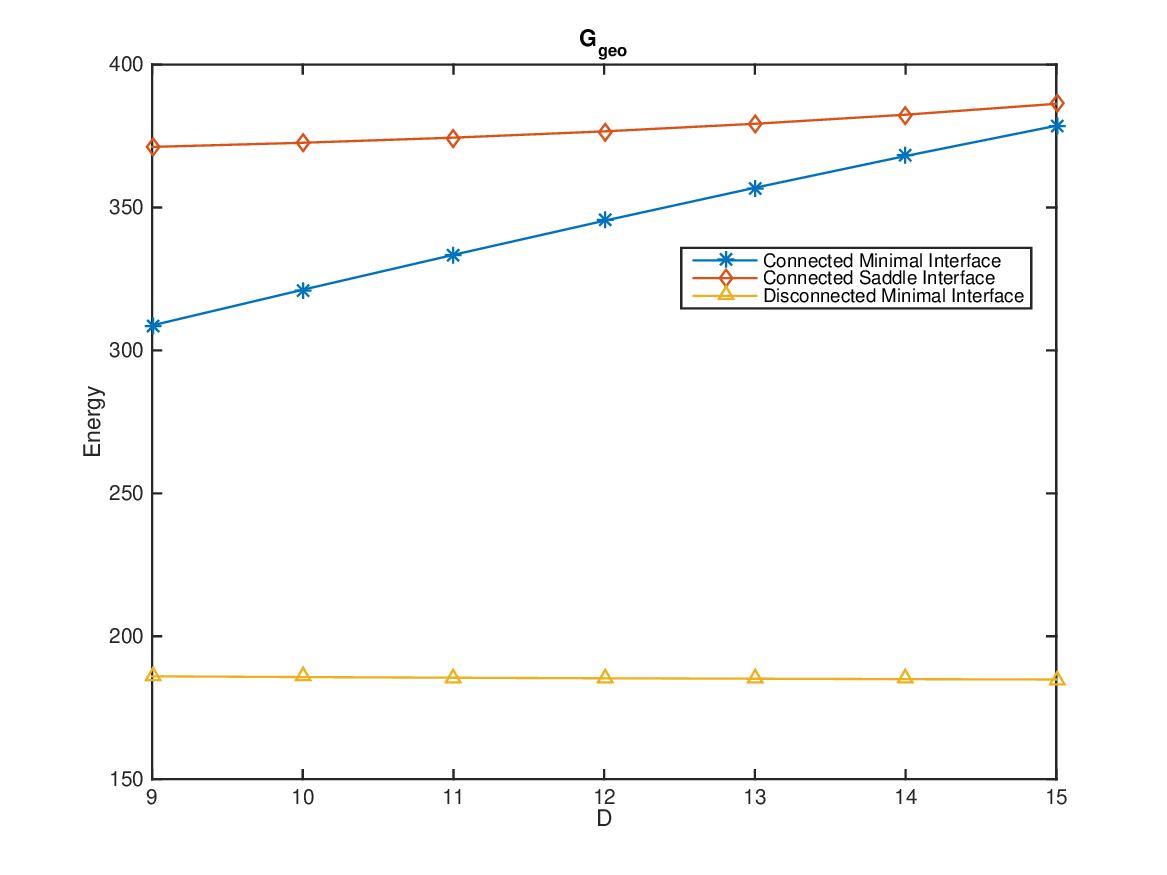}
\end{minipage}
\begin{minipage}[c]{7.26cm}
\centering
\includegraphics[scale=0.4]{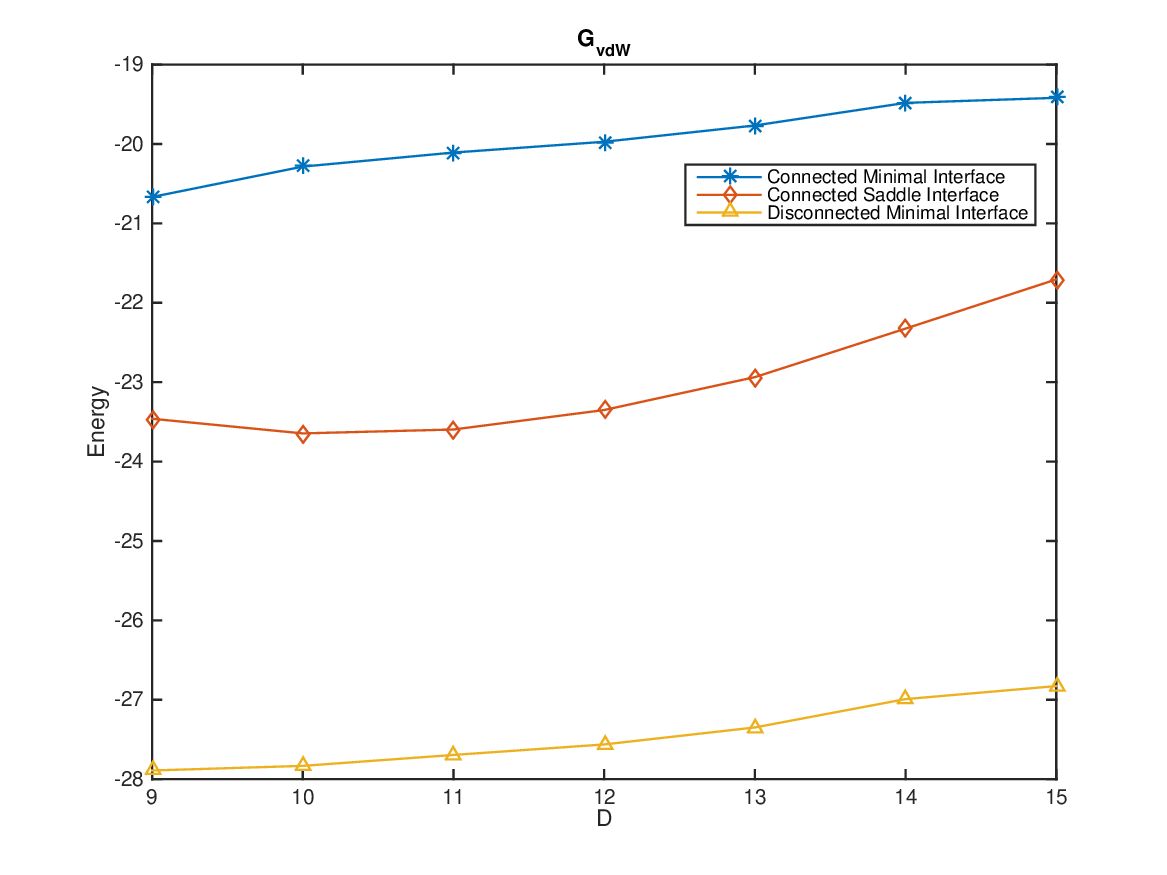}
\end{minipage}
\caption{\label{Fig:G_not_tot_2}The varying trend of $G_{\text{geo}}$ with respect to $D$ (left); The varying trend of $G_{\text{vdW}}$ with respect to $D$ (right).}
\end{figure}

\section*{Acknowledgements}

\noindent
The author highly appreciates the supervision of Prof.~Dr.~Shenggao Zhou. The author also thanks his colleague student Cheng Tao for many helpful discussions. This paper is part of the author's bachelor thesis.

\end{document}